\PassOptionsToPackage{unicode}{hyperref}
\PassOptionsToPackage{hyphens}{url}
\PassOptionsToPackage{dvipsnames,svgnames,x11names}{xcolor}
\documentclass[
  12pt]{article}

\usepackage{amsmath,amssymb}
\usepackage{iftex}
\ifPDFTeX
  \usepackage[T1]{fontenc}
  \usepackage[utf8]{inputenc}
  \usepackage{textcomp} 
\else 
  \usepackage{unicode-math}
  \defaultfontfeatures{Scale=MatchLowercase}
  \defaultfontfeatures[\rmfamily]{Ligatures=TeX,Scale=1}
\fi
\usepackage{lmodern}
\ifPDFTeX\else  
\fi
\IfFileExists{upquote.sty}{\usepackage{upquote}}{}
\IfFileExists{microtype.sty}{
  \usepackage[]{microtype}
  \UseMicrotypeSet[protrusion]{basicmath} 
}{}
\makeatletter
\@ifundefined{KOMAClassName}{
  \IfFileExists{parskip.sty}{%
    \usepackage{parskip}
  }{
    \setlength{\parindent}{0pt}
    \setlength{\parskip}{6pt plus 2pt minus 1pt}}
}{
  \KOMAoptions{parskip=half}}
\makeatother
\usepackage{xcolor}
\makeatletter
\ifx\paragraph\undefined\else
  \let\oldparagraph\paragraph
  \renewcommand{\paragraph}{
    \@ifstar
      \xxxParagraphStar
      \xxxParagraphNoStar
  }
  \newcommand{\xxxParagraphStar}[1]{\oldparagraph*{#1}\mbox{}}
  \newcommand{\xxxParagraphNoStar}[1]{\oldparagraph{#1}\mbox{}}
\fi
\ifx\subparagraph\undefined\else
  \let\oldsubparagraph\subparagraph
  \renewcommand{\subparagraph}{
    \@ifstar
      \xxxSubParagraphStar
      \xxxSubParagraphNoStar
  }
  \newcommand{\xxxSubParagraphStar}[1]{\oldsubparagraph*{#1}\mbox{}}
  \newcommand{\xxxSubParagraphNoStar}[1]{\oldsubparagraph{#1}\mbox{}}
\fi
\makeatother

\usepackage{longtable,booktabs,array}
\usepackage{calc} 
\usepackage{etoolbox}
\makeatletter
\patchcmd\longtable{\par}{\if@noskipsec\mbox{}\fi\par}{}{}
\makeatother
\IfFileExists{footnotehyper.sty}{\usepackage{footnotehyper}}{\usepackage{footnote}}
\makesavenoteenv{longtable}
\usepackage{graphicx}
\makeatletter
\def\maxwidth{\ifdim\Gin@nat@width>\linewidth\linewidth\else\Gin@nat@width\fi}
\def\maxheight{\ifdim\Gin@nat@height>\textheight\textheight\else\Gin@nat@height\fi}
\makeatother
\setkeys{Gin}{width=\maxwidth,height=\maxheight,keepaspectratio}
\makeatletter
\def\fps@figure{htbp}
\makeatother

\makeatletter
\@ifpackageloaded{caption}{}{\usepackage{caption}}
\AtBeginDocument{%
\ifdefined\contentsname
  \renewcommand*\contentsname{Table of contents}
\else
  \newcommand\contentsname{Table of contents}
\fi
\ifdefined\listfigurename
  \renewcommand*\listfigurename{List of Figures}
\else
  \newcommand\listfigurename{List of Figures}
\fi
\ifdefined\listtablename
  \renewcommand*\listtablename{List of Tables}
\else
  \newcommand\listtablename{List of Tables}
\fi
\ifdefined\figurename
  \renewcommand*\figurename{Figure}
\else
  \newcommand\figurename{Figure}
\fi
\ifdefined\tablename
  \renewcommand*\tablename{Table}
\else
  \newcommand\tablename{Table}
\fi
}
\@ifpackageloaded{float}{}{\usepackage{float}}
\floatstyle{ruled}
\@ifundefined{c@chapter}{\newfloat{codelisting}{h}{lop}}{\newfloat{codelisting}{h}{lop}[chapter]}
\floatname{codelisting}{Listing}

\makeatother
\makeatletter
\@ifpackageloaded{caption}{}{\usepackage{caption}}
\@ifpackageloaded{subcaption}{}{\usepackage{subcaption}}
\makeatother

\ifLuaTeX
  \usepackage{selnolig}  
\fi
\usepackage[]{natbib}
\usepackage{bookmark}

\IfFileExists{xurl.sty}{\usepackage{xurl}}{} 
\hypersetup{
  colorlinks=true,
  linkcolor={blue},
  filecolor={Maroon},
  citecolor={Blue},
  urlcolor={Blue}
}

\usepackage[most]{tcolorbox}

\newtcolorbox{examplebox}[1][]{
  enhanced,
  breakable,
  colback=white,
  colframe=black,
  boxrule=0.6pt,
  arc=1pt,
  left=6pt,
  right=6pt,
  top=6pt,
  bottom=6pt,
  title=#1,
  fonttitle=\bfseries
}

\usepackage{tikz}
\usepackage{url}
\usepackage{booktabs}
\usepackage{amsfonts}
\usepackage{nicefrac}
\usepackage{array}
\usepackage{microtype}
\usepackage{xcolor}
\usepackage{graphicx}
\usepackage{amsmath,amssymb,amsthm}
\usepackage{algorithm}
\usepackage{algpseudocode}
\usepackage{comment}
\usepackage{multirow}
\usepackage{svg}
\usepackage{float}
\usepackage{hyperref}
\usepackage{listings}
\usetikzlibrary{arrows.meta, positioning, calc}

\newcommand{\anon}{1}

\begin{document}

\def\spacingset#1{\renewcommand{\baselinestretch}%
{#1}\small\normalsize} \spacingset{1}


\if1\anon
{
  \title{\bf AI4BayesCode: From Natural Language Descriptions to Validated Modular Stateful Bayesian Samplers}
\author{
Jungang Zou 
\hspace{.4cm}
Alex Ziyu Jiang
\hspace{.4cm}
Qixuan Chen 
\\
Department of Biostatistics, Columbia University
}
  \maketitle
} \fi

\if0\anon
{
  \bigskip
  \bigskip
  \bigskip
  \begin{center}
    {\LARGE\bf AI4BayesCode: From Natural Language Descriptions to Validated Modular Stateful Bayesian Samplers}
\end{center}
  \medskip
} \fi

\bigskip
\begin{abstract}
Coding and computation remain major bottlenecks in Markov chain Monte Carlo (MCMC) workflows, especially as modern sampling algorithms have become increasingly complex and existing probabilistic programming systems remain limited in model support, extensibility, and composability. We introduce \textbf{AI4BayesCode}, an extensible LLM-driven system that translates natural-language Bayesian model descriptions into runnable, validated MCMC samplers. To improve reliability, AI4BayesCode adopts a modular design that decomposes models into modular sampling blocks and maps each block to a built-in sampling component, reducing the need to implement complex sampling algorithms from scratch. Reliability is further improved through pre-generation validation of model specifications and post-generation validation of generated sampler code. AI4BayesCode also introduces a novel recursively stateful coding paradigm for MCMC, allowing modular sampling components, potentially developed by different contributors, to be composed coherently within larger MCMC procedures. We develop a benchmark suite to evaluate AI4BayesCode for sampler-generation. Experiments show that AI4BayesCode can implement a wide range of Bayesian models from natural-language descriptions alone. As an open-ended system, its capability can continue to expand with improvements in the underlying AI agent and the addition of new built-in blocks. The software is available at \url{https://ai4bayescode.com/}.
\end{abstract}

\noindent%
{\it Keywords:} AI agentic programming; AI-generated code validation; Bayesian computation; Large language models; Markov chain Monte Carlo; 
\vfill

\newpage
\spacingset{1.8} 

\section{Introduction}\label{sec-intro}
Markov chain Monte Carlo (MCMC) has long been a dominant tool for Bayesian inference, with methods ranging from Gibbs samplers \citep{Geman1984} for conjugate models and Metropolis-Hastings \citep{Metropolis1953, Hastings1970} for non-conjugate settings to  advanced gradient-based algorithms such as Hamiltonian Monte Carlo \citep{Betancourt2013, Brooks_2011}, the Metropolis-adjusted Langevin algorithm \citep{Roberts1996}, and the no-U-turn sampler (NUTS; \citet{Hoffman2014}), which exploit log-posterior gradients for more efficient exploration. 
However, implementing these methods can be challenging. Modern samplers often involve complex components such as leapfrog integrators, dual-averaging step-size adaptation, recursive tree doubling, and constraint-transform Jacobians. Writing, debugging, and validating such code require deep expertise in Bayesian statistics and scientific computing, limiting the range of models that researchers lacking expertise in these areas can practically explore. 

Probabilistic programming language systems such as \texttt{Stan} \citep{Carpenter2017}, \texttt{BUGS} \citep{Lunn2000}, \texttt{JAGS} \citep{Plummer2003}, \texttt{PyMC} \citep{pymc2023}, \texttt{NumPyro} \citep{Phan2019}, and \texttt{Turing.jl} \citep{Fjelde2025} have narrowed this gap by providing declarative modeling languages with automated sampler construction. However,  they still require users to write modeling code and face several structural limitations. First, support for certain model classes is limited: for example, \texttt{BUGS}/\texttt{JAGS} lack gradient-based samplers, while \texttt{Stan} offers only limited support for discrete parameters. Second, each platform defines its own syntax and type system, making it difficult to combine components across platforms or embed fitted components into a larger MCMC model. Third, the incorporation of new models and algorithms is often slow; for instance, Bayesian additive regression trees (BART; \citet{Chipman2010}) were integrated only long after their original formulation. These limitations highlight the need for a more flexible and extensible framework that reduces the burden of manual coding, supports a broader class of models, and enables rapid integration of new methods.

Large language models (LLMs) have been applied to the Bayesian modeling process. Narrow applications include prior elicitation \citep{Huang2025,Riegler2025,Capstick2025,Li2025}, concept selection \citep{Feng2025}, and natural-language translation of priors and likelihoods for regression models \citep{Huang2025a}. At the workflow level, LLM-based agents automate model discovery by iteratively proposing, refining, and evaluating model structures using predictive and diagnostic feedback \citep{Sun2026,Duerr2026}. These systems rely on existing probabilistic programming languages as the execution substrate, limiting their scope to models that can be expressed within those languages. More broadly, AI-assisted coding paradigms such as agentic programming \citep{Wang2025} and ``vibe coding'' \citep{Ge2025} enable direct code generation from natural-language specifications, but the resulting code is often prone to subtle, hard-to-detect bugs \citep{Anand2024}. A separate and largely unaddressed question is MCMC sampler construction: while LLMs can assist with simple samplers, generating correct implementations of complex algorithms such as NUTS or BART remains difficult, and the resulting code is rarely validated.

A fundamental limitation of both probabilistic programming languages and LLM-generated MCMC code is that samplers are typically treated as monolithic objects. For example, the \texttt{R} package \texttt{BART} (version 2.9.10) \citep{Sparapani2021} executes its sampler as a single call; embedding it within a larger MCMC requires reinitializing the entire ensemble at each iteration, discarding all learned tree structure. What is missing is a \textit{stateful} interface in which each sampling component, at every iteration, receives the latest values of all other components as input and updates its own parameters based on both this external input and its internally maintained state from the previous iteration. For instance, a BART component embedded in a larger model should be able to retain its current tree ensemble, accept updated inputs (such as imputed predictors or latent working response variables), and then update its tree ensemble accordingly without reinitialization. Such a paradigm would allow independently developed components---gradient-based samplers, conjugate Gibbs updates, and tree ensembles---to be composed within a single unified iterative framework.

To address these issues, we introduce \textbf{AI4BayesCode}, an LLM-driven system that translates natural-language Bayesian model descriptions into validated, modular, and stateful MCMC samplers. The system first parses a natural-language  description into a structured model formulation, then decomposes it into modular blocks, reducing complex samplers into simpler components that are easier to implement and validate. Each block is assigned to a built-in sampling module, avoiding the need to implement complex algorithms from scratch. AI4BayesCode adopts a recursively stateful paradigm: each modular block is stateful, and the full sampler composed from these blocks is itself stateful. This allows updated external quantities to be propagated from the outer sampler to inner blocks, while each block retains its current internal state and continues sampling without reinitialization within a larger MCMC procedure. To improve reliability, this generation process is coupled with a two-stage validation framework: pre-generation validation ensures the parsed model matches the user’s intent, and post-generation validation evaluates the sampler code through syntax, semantic, and runtime diagnostics. The output is a validated sampler with unified stateful functions and bindings for multiple front-end languages, including \texttt{C++}, \texttt{R}, and \texttt{Python}. We provide details on system design and systematic validation in Sections~\ref{sec:design} and~\ref{sec:validation}, respectively. We then illustrate the approach by constructing a sampler for Dirac spike-and-slab regression in Section~\ref{sec:spike_slab}, followed by three experiments evaluating the performance of the AI4BayesCode system in Section~\ref{s:experiments}.

Our contributions are as follows. We introduce AI4BayesCode, to our knowledge the first LLM-driven system for constructing validated MCMC samplers directly from natural-language Bayesian model descriptions. Central to this system is a novel recursively stateful coding paradigm that enables modular sampling components, potentially developed by different contributors, to be composed coherently within larger MCMC procedures. To ensure reliability, we develop a two-stage validation framework that systematically detects and reduces errors in AI-generated sampler implementation. Finally, we construct a benchmark suite to evaluate this system, consisting of natural-language model descriptions, data-generation scripts, and reference implementations whenever available.

\section{System design}
\label{sec:design}

\begin{figure}[h]
\begin{center}
\includegraphics[width=\textwidth,height=0.35\textheight]{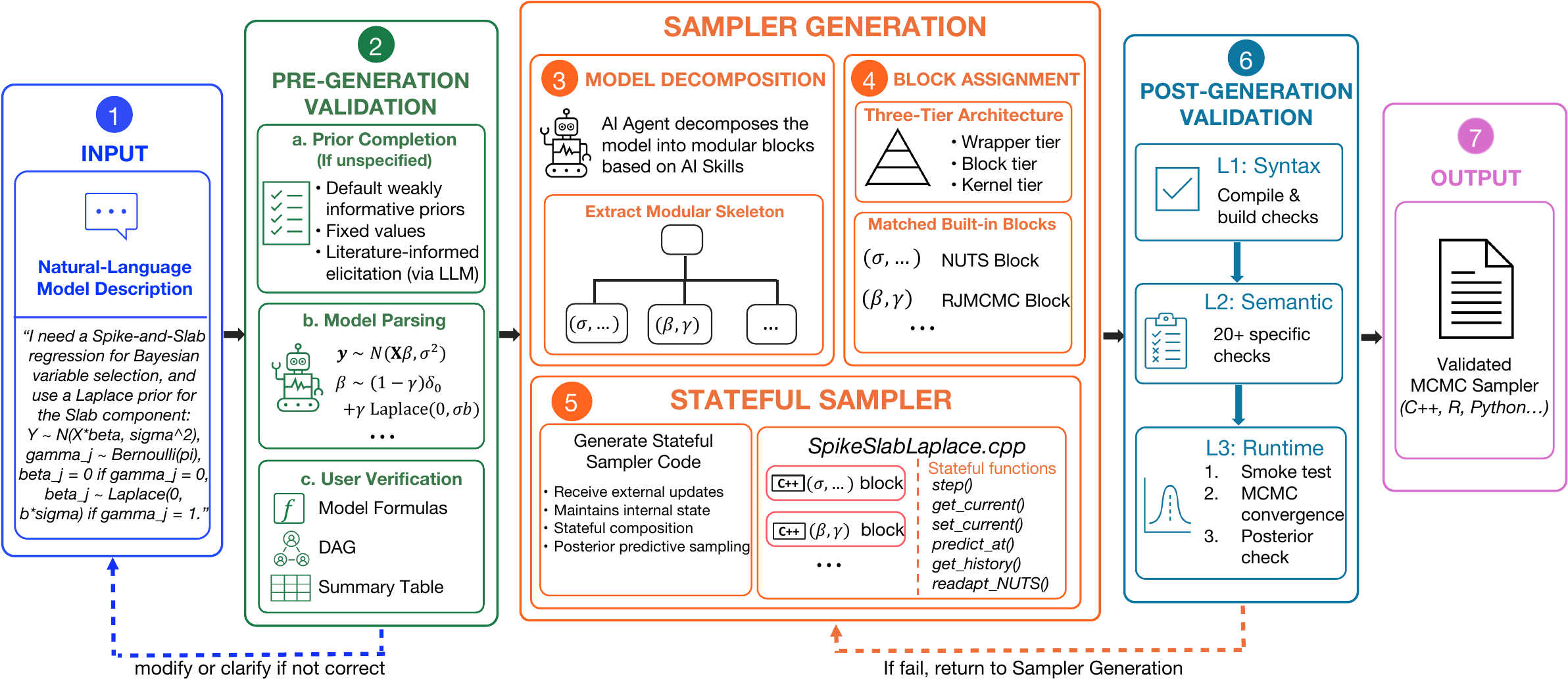}
\end{center}
\caption{\em Overview of AI4BayesCode using a Spike-and-Slab regression model as an example. The system translates a natural-language Bayesian model description into a validated MCMC sampler through pre-generation validation, modular block decomposition and assignment, stateful sampler generation, and post-generation validation. DAG: directed acyclic graph. NUTS: no-U-turn sampler. RJMCMC: reversible-jump MCMC.}
\label{fig:flowchart}
\end{figure}

AI4BayesCode is an AI-powered system that translates natural-language Bayesian model descriptions into validated, modular, and stateful MCMC samplers. {Rather than training a specialized LLM, it is a system built on existing LLM code-generating agents, together with AI Skills, reference examples, built-in sampling blocks, and validation checks.} Figure~\ref{fig:flowchart} summarizes the overall workflow. After the user enters a Bayesian model description, an AI agent first parses the natural-language input into a structured model specification, including the likelihoods, priors, parameter constraints, and dependencies. This section focuses on two key design components that transform the model specification into sampler code. The first is a modular code generation design that decomposes the model into modular sampling blocks (Steps 3 and 4 in Figure~\ref{fig:flowchart}). The second is a stateful sampler paradigm that specifies how these blocks retain internal states, accept updated external inputs, and compose recursively within larger MCMC workflows (Step 5). The validation framework, which verifies both the user-specified model and the generated sampler code (Step 2 and 6), is described separately in Section~\ref{sec:validation}.

\subsection{Modular system architecture}

\subsubsection{Design principle}

Modularity is the core design principle of the AI4BayesCode generator, where samplers are constructed from pre-validated update kernels rather than a full joint posterior. This design is motivated by two key challenges in AI-generated samplers. First, constructing a full joint sampler requires the LLM to correctly specify all the likelihoods, priors, constraints, and transformations in a single joint posterior; any omission or incorrect parameterizations can silently corrupt the resulting chain. Second, complex algorithms such as NUTS and RJMCMC kernels are difficult for an LLM to implement reliably from scratch. To address these issues, AI4BayesCode partitions the full model into modular blocks, each corresponding to one or a group of parameters, to reduce the complexity of updates and allows each component to be implemented and validated separately. From a probabilistic perspective, each update kernel depends only on a subset of model components rather than the full posterior distribution. Specifically, let $\theta$ denote the set of parameters in one modular block, $\mathcal D$ denote all observed data, $\Theta$ denote all parameters in this model, and $\Theta_{-\theta}$ denote all parameters other than $\theta$. By conditional independence, the update for the $\theta$ block can be reduced from the full joint posterior $\pi(\theta, \Theta_{-\theta} \mid \mathcal D)$ to a modular conditional target:
\begin{equation}
\pi(\theta \mid \Theta_{-\theta}, \mathcal D)
 \propto
p(\mathcal D \mid \theta, \Theta_{-\theta})\,\pi(\theta \mid \Theta_{-\theta}) \propto
p(\mathcal Z_\theta \mid \theta, \mathcal C_\theta)\,\pi(\theta \mid \mathcal H_\theta) \propto
\pi(\theta \mid \mathcal Z_\theta, \mathcal C_\theta, \mathcal H_\theta),
\end{equation}
where $\mathcal Z_\theta$ denotes the observed or latent quantities whose conditional distribution involves $\theta$ and other conditioning variables $\mathcal C_\theta$, and $\pi(\theta \mid \mathcal H_\theta)$ is the prior of $\theta$ indexed by $\mathcal H_\theta$.

\begin{examplebox}[Example 1. Modular block decomposition in Gaussian linear regression]
Consider the Gaussian linear regression model
\[
y \sim \mathcal{N}(X\beta, \sigma^2 I), \qquad
\beta \sim \mathcal{N}(\mu, \eta^2 I), \qquad
\sigma \sim \mathcal{C}^+(0, \tau_\sigma),
\]
where $\mathcal{C}^+(0, \tau_\sigma)$ denotes a half-Cauchy prior with scale $\tau_\sigma$, and $\mu$ and $\eta^2$ are hyperparameters. 

For the block $\theta = \{\sigma\}$, the modular conditional target is $\pi(\sigma \mid y, X\beta, \tau_\sigma)
\propto
p(y \mid \sigma, X\beta)\,\pi(\sigma \mid \tau_\sigma)$, which does not involve $\mu$ or $\eta^2$ once $\beta$ is conditioned on. Thus, the relevant quantities for this block are $\mathcal{Z}_\theta = \{y\}$, $\mathcal{C}_\theta = \{X\beta\}$ and $\mathcal{H}_\theta = \{\tau_\sigma\}$. For the block $\theta = \{\beta\}$, similarly, the corresponding quantities are $\mathcal{Z}_\theta = \{y\}$, $\mathcal{C}_\theta = \{X, \sigma\}$, and $\mathcal{H}_\theta = \{\mu, \eta^2\}$.

\end{examplebox}


\subsubsection{Three-tier architecture} 
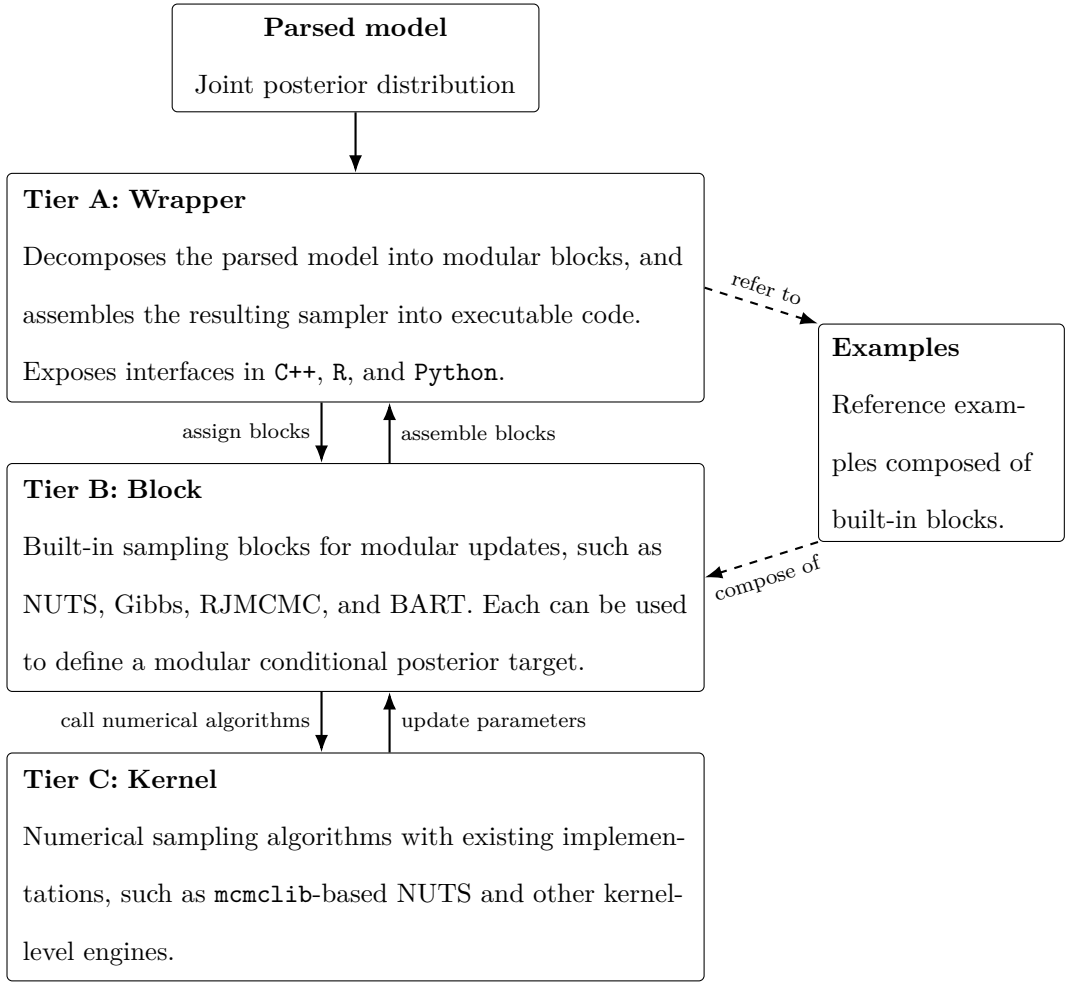
\begin{figure}[h]
\centering
\begin{tikzpicture}[
  >=Latex,
  node distance=0.8cm,
  tier/.style={
    draw,
    rounded corners=2pt,
    align=left,
    text width=8.8cm,
    inner sep=6pt,
    font=\footnotesize
  },
  boxsmall/.style={
    draw,
    rounded corners=2pt,
    align=center,
    text width=4.5cm,
    inner sep=5pt,
    font=\footnotesize
  },
  exbox/.style={
    draw,
    rounded corners=2pt,
    align=left,
    text width=2.9cm,
    inner sep=5pt,
    font=\footnotesize
  },
  lab/.style={font=\scriptsize}
]

\node[boxsmall] (D) {\textbf{Parsed model}\\Joint posterior distribution};

\node[tier, below=of D] (A) {\textbf{Tier A: Wrapper}\\
Decomposes the parsed model into modular blocks, and assembles the resulting sampler into executable code.\\
Exposes interfaces in \texttt{C++}, \texttt{R}, and \texttt{Python}.};

\node[tier, below=of A] (B) {\textbf{Tier B: Block}\\
Built-in sampling blocks for modular updates, such as NUTS, Gibbs, RJMCMC, and BART. Each can be used to define a modular conditional posterior target.};

\node[tier, below=of B] (C) {\textbf{Tier C: Kernel}\\
Numerical sampling algorithms with existing implementations, such as \texttt{mcmclib}-based NUTS and other kernel-level engines.};

\node[exbox, anchor=west] (E) at ($(A.south east)!0.5!(B.north east)+(1.5cm,0)$) {
\textbf{Examples}\\
Reference examples composed of built-in blocks.
};

\draw[->, thick] (D.south) -- (A.north);

\draw[->, thick]
  ([xshift=-0.45cm]A.south) --
  node[left, lab]{assign blocks}
  ([xshift=-0.45cm]B.north);

\draw[->, thick]
  ([xshift=-0.45cm]B.south) --
  node[left, lab]{call numerical algorithms}
  ([xshift=-0.45cm]C.north);

\draw[->, thick]
  ([xshift=0.45cm]C.north) --
  node[right, lab]{update parameters}
  ([xshift=0.45cm]B.south);

\draw[->, thick]
  ([xshift=0.45cm]B.north) --
  node[right, lab]{assemble blocks}
  ([xshift=0.45cm]A.south);

\draw[->, dashed, thick] (A.east) -- node[above, sloped, lab]{refer to} (E.north west);
\draw[->, dashed, thick] (E.south west) -- node[below, sloped, lab]{compose of} (B.east);

\end{tikzpicture}
\caption{\em Three-tier architecture of AI4BayesCode.}
\label{fig:three_tier_architecture}
\end{figure}

AI4BayesCode addresses the difficulty of reliably implementing complex algorithms from scratch using an LLM through a three-tier architecture, in which modular blocks are updated using built-in sampling components rather than code generated from scratch. The architecture consists of three layers (Figure~\ref{fig:three_tier_architecture}): a \textit{wrapper} tier, a \textit{block} tier, and a \textit{kernel} tier. The wrapper first decomposes the parsed model into modular blocks. At the block tier, the AI agent uses system-provided reference examples to assign each modular block to an appropriate built-in sampling block (e.g., NUTS, RJMCMC, Gibbs, or BART block) for model-specific updates. Rather than requiring the AI agent to implement sampling algorithms from scratch, the system provides existing numerical sampling algorithms in the kernel tier (e.g., NUTS built on \texttt{mcmclib} \citep{OHara2023}). The block tier calls the kernel tier to perform sampling updates, and the blocks are updated sequentially to complete one MCMC iteration. Finally, the wrapper assembles these modular sampling blocks into executable code. The system and the whole sampler are implemented in \texttt{C++} for computational efficiency, with the wrapper layer providing direct \texttt{C++} support along with interfaces to \texttt{R} and \texttt{Python}.

A key feature of this architecture is the separation between model-specific modular targets and generic sampling algorithms. The block tier is responsible for the former: for each modular block $\theta$, it specifies the conditional target $\pi(\theta \mid \mathcal Z_\theta, \mathcal C_\theta, \mathcal H_\theta)$ with $\mathcal Z_\theta$, $\mathcal C_\theta$, and $\mathcal H_\theta$ and their relationships. The kernel tier is responsible for the latter by providing generic transition mechanisms, such as NUTS and BART-related algorithms, that can be shared across multiple blocks. Under this separation, the AI agent does not need to implement complex sampling algorithms from scratch. Instead, it only needs to construct the modular conditional posterior and assign it to an appropriate built-in block connected to a pre-built sampling kernel. This design improves system reliability by largely restricting AI-generated code to model-specific target construction rather than the implementation of sampling algorithms.

\begin{examplebox}[Example 2. Three-tier update in Gaussian linear regression]
Continuing Example 1, the wrapper decomposes the model and assigns $\sigma$ and $\beta$ to two separate modular blocks, each associated with its own conditional target: \[ \pi(\sigma \mid y, X\beta, \tau_\sigma) \qquad \text{and} \qquad \pi(\beta \mid y, X, \sigma, \mu, \eta^2). \] By default, AI4BayesCode uses the NUTS block for continuous parameters and thus assigns both modular blocks to NUTS samplers. The same kernel-tier NUTS algorithm, implemented using \texttt{mcmclib}, is used to perform sampling updates in each block, although the two blocks require different block-specific specifications. The blocks are then updated sequentially to complete one MCMC iteration.

We note that this decomposition is presented only for illustrative purposes. In practice, we typically update the continuous parameter block $\theta=\{\sigma,\beta\}$ jointly within a single NUTS transition. This joint update can improve sampling efficiency when the continuous parameters are coupled.


\end{examplebox}

\subsubsection{Model decomposition and Skill-guided block selection}
An essential component of the system is the provision of AI Skills and block-level reference examples. Given the parsed model formulation, these built-in AI Skills and block-tier reference examples guide the code generator to decompose the model into modular blocks and assign each modular block to an appropriate built-in block. They also define the priority of block selection. For example, in the Spike-and-Slab model illustrated in Figure~\ref{fig:flowchart}, the reference Skills assign $(\beta,\gamma)$ to a joint RJMCMC block rather than separate updates, recognizing the strong coupling between the parameters and the need for a joint trans-dimensional sampler. Conjugate Gibbs updates are assigned the lowest priority in AI4BayesCode and are used only when necessary, because they require the AI to correctly identify conjugacy and derive the full conditional under the appropriate parameterization, which may introduce difficult-to-detect errors.

\subsubsection{Extensible system}
As an open-ended system, AI4BayesCode can be extended at two levels. At the block tier, users can develop custom built-in sampling blocks and reference examples through AI Skills to support domain-specific models. At the kernel tier, new computational algorithms can be incorporated as long as they expose the stateful functions (see Section~\ref{sec:stateful}) required by the block-tier sampler. For example, the system supports BART-family models by adapting the \texttt{BART} package \citep{Sparapani2021} with stateful kernel-tier functions that can be called by the block-tier BART block. A list of built-in blocks in the current version of AI4BayesCode is provided in Supplementary Materials S1.

\subsection{Stateful sampler paradigm} \label{sec:stateful}

\subsubsection{Design principle}
Statefulness is central to the sampler code generated by AI4BayesCode. In a conventional non-stateful design, a sampler with parameters $\Theta$ is run on observed data $\mathcal D$, targeting $\pi(\Theta \mid \mathcal D)$. Let $(t)$ denote the $t$-th MCMC iteration. A non-stateful design runs a chain of the form 
\begin{equation}\label{eq:nonstateful}
\dots\longrightarrow\; (\Theta^{(t)}, \mathcal D)\;\longrightarrow\;(\Theta^{(t+1)}, \mathcal D)\longrightarrow\; \dots.
\end{equation}
When embedded within a larger MCMC procedure, however, $\mathcal D$ may change across iterations, for example through updated working responses or imputed predictors, requiring the sampler to discard the current state of $\Theta$ and be reinitialized for the new target. 
By contrast, 
a stateful design allows the conditioning quantities to be updated while retaining the current state of $\Theta$:
\begin{equation}\label{eq:stateful}
\dots\longrightarrow\; (\Theta^{(t)}, \mathcal D^{(t)})\;\longrightarrow\;(\Theta^{(t)}, \mathcal D^{(t+1)})\;\longrightarrow\;(\Theta^{(t+1)}, \mathcal D^{(t+1)})\longrightarrow\; \dots.
\end{equation}
Thus, a stateful sampler can incorporate updated quantities $ \mathcal D^{(t+1)}$ from other blocks and continue sampling $\Theta^{(t+1)}$ without reinitialization. Within the outer MCMC, the sampler for $\Theta$, together with the blocks updating $\mathcal D$, jointly targets the corresponding conditional posterior $\pi(\Theta,\mathcal D \mid \cdot)$. An example of the stateful design with updated working responses for meta-regression is provided in~\ref{app:meta_regression}.

\subsubsection{Stateful functions}
\label{sec:stateful_funs}
To implement the stateful update, three core functions are necessary for each stateful sampler: \texttt{set\_current()}, which receives updated conditioning quantities $\mathcal D^{(t+1)}$ from external blocks; \texttt{step()}, which updates the sampler state from $\Theta^{(t)}$ to $\Theta^{(t+1)}$ given $\mathcal D^{(t+1)}$; and \texttt{get\_current()}, which passes the current state to other blocks. Although the sampler operates in a stateful manner, AI4BayesCode also supports recording the full sampling trajectory of parameters, which can be accessed through the \texttt{get\_history()} function. 

\subsubsection{Recursive composition}

\begin{algorithm}[h] \caption{\texttt{step()} function of a stateful sampler} \label{alg:stateful_step} \begin{algorithmic}[1] 
\Require Current sampler state $\Theta^{(t)}$, shared pool $\mathcal P$ 
\For{each modular block $\theta_k \subseteq \Theta$} 
\State Read required quantities $(\mathcal Z_{\theta_k}^{(t)}, \mathcal C_{\theta_k}^{(t)}, \mathcal H_{\theta_k}^{(t)})$ from $\mathcal P$ 
\State $\theta_k:\texttt{set\_current}(\mathcal Z^{(t)}_{\theta_k}, \mathcal C^{(t)}_{\theta_k}, \mathcal H^{(t)}_{\theta_k})$ 
\State $\theta_k:\texttt{step}()$ \Comment{Recursively calls inner blocks if this block is composite} 
\State $\theta_k^{(t+1)} \gets \theta_k:\texttt{get\_current}()$ 
\State Update $\mathcal P$ with $\theta_k^{(t+1)}$ \EndFor \State \Return $\Theta^{(t+1)}$ \end{algorithmic} \end{algorithm}

AI4BayesCode implements both modular blocks and the full generated sampler in a recursively stateful manner.  
The full sampler is organized hierarchically: a sampler may contain multiple blocks, and higher-level composite blocks may themselves contain lower-level blocks. Each block maintains its own internal state, while higher-level samplers coordinate updates across their constituent blocks. During one MCMC iteration, the \texttt{step()} function of the full sampler calls the \texttt{step()} functions of its constituent blocks, which may recursively call the \texttt{step()} functions of lower-level blocks until the lowest-level blocks are reached, where the kernel-tier algorithms are executed. 

This recursive update structure would require each block to repeatedly call \texttt{get\_current()} from many other blocks to retrieve conditioning quantities. As the number of blocks grows, these cross-block calls become increasingly difficult for AI-generated code to specify correctly and can lead to missing or outdated conditioning quantities. To support this hierarchical orchestration, each composite block or sampler maintains a \textit{shared pool} that contains the block-specific quantities at iteration $t$. For a modular inner block corresponding to $\theta$, its \texttt{set\_current()} function reads the required quantities $\mathcal Z_\theta^{(t)}$, $\mathcal C_\theta^{(t)}$, and $\mathcal H_\theta^{(t)}$ from the shared pool, its \texttt{step()} function updates the state from $\theta^{(t)}$ to $\theta^{(t+1)}$, and its \texttt{get\_current()} function writes $\theta^{(t+1)}$ back to the shared pool. Algorithm~\ref{alg:stateful_step} summarizes this stateful update using the \texttt{step()} function in AI4BayesCode samplers. This recursive stateful design allows independently developed components to be embedded and updated systematically within a larger MCMC procedure, thereby enabling AI4BayesCode to build highly complex Bayesian samplers from modular or separately developed blocks.


\subsubsection{Posterior predictive sampling}
AI4BayesCode supports posterior predictive sampling via \texttt{predict\_at()}, covering both unconditional (e.g., density estimation) and conditional (e.g., regression with new values of covariates) settings. For each sampler, the system builds an internal prediction directed acyclic graph (DAG) to identify the prediction setting and determine which downstream variables to predict. When the sampling trajectory history is retained,  \texttt{predict\_at()} returns predictions over the full posterior trajectory; otherwise, it returns predictions for the current draw only. See~\ref{app:meta_regression} for a continued illustration using the meta-regression example introduced earlier.

\section{Systematic validation}
\label{sec:validation}

AI4BayesCode employs a two-stage validation framework to improve the reliability of generated samplers for complex models by detecting and resolving errors in model specification, sampler implementation, and posterior computation. The framework includes a pre-generation validation step to verify that the intended model is correctly captured, and a post-generation step to assess the sampler code through syntax, semantic, and runtime diagnostics.

\subsection{Pre-generation validation}
The pre-generation validation step is an interactive phase conducted before code generation to reduce errors in model formulation (Step 2 in Figure~\ref{fig:flowchart}). The first component is prior completion (Step 2a), where, if priors are unspecified, AI4BayesCode presents options including fixed values, default non- or weakly informative priors, and literature-informed priors obtained via LLM-assisted search \citep{Riegler2025}. The second component is user verification (Step 2c), a user-in-the-loop process to improve reliability when model specifications are ambiguous or underspecified \citep{Ge2025,Fakhoury2024}. The system provides three structured summaries for user review: (1) model formulas, which make the parsed likelihood and prior specification explicit; (2) a DAG of model dependencies, which visualizes the dependency structure among model components; and (3) a summary table, which gives an overview of the model specification. When ambiguities remain (e.g., different parameterizations of the same distribution), the system explicitly prompts the user for clarification.

\subsection{Post-generation validation}
Post-generation validation is conducted after code generation to assess the sampler (Step 6 in Figure~\ref{fig:flowchart}). It proceeds using a three-level pipeline: code compilation, implementation checks, and posterior validation. If the sampler fails at any level, the LLM revises the code based on accumulated feedback and attempts until a pre-specified limit \(M\) is reached. 
The first two levels cover syntax and semantics. Syntax validation ensures the code is syntactically valid and can be compiled successfully. Semantic validation comprises more than twenty checks targeting common MCMC implementation errors that may compile and run but yield incorrect posteriors, such as misparameterization, dimension mismatch, inconsistent dependencies, and invalid update logic (see Supplementary Materials S2). The third level is runtime validation, which evaluates the sampler on a simulated dataset generated under the target model. It includes a short smoke test, followed by two-chain rank-normalized \(\hat R\) and effective sample size (ESS) diagnostics \citep{Vehtari2021}, and posterior checks using predictive summaries and PSIS-LOO \citep{Vehtari2024} when available (see Supplementary Materials S3). These checks aim to detect failures such as numerical instability, poor mixing behavior, or posterior miscalibration.

\section{Usage example}
\label{sec:spike_slab}

\begin{figure}[p]
\centering
\includegraphics[width=\textwidth,keepaspectratio]{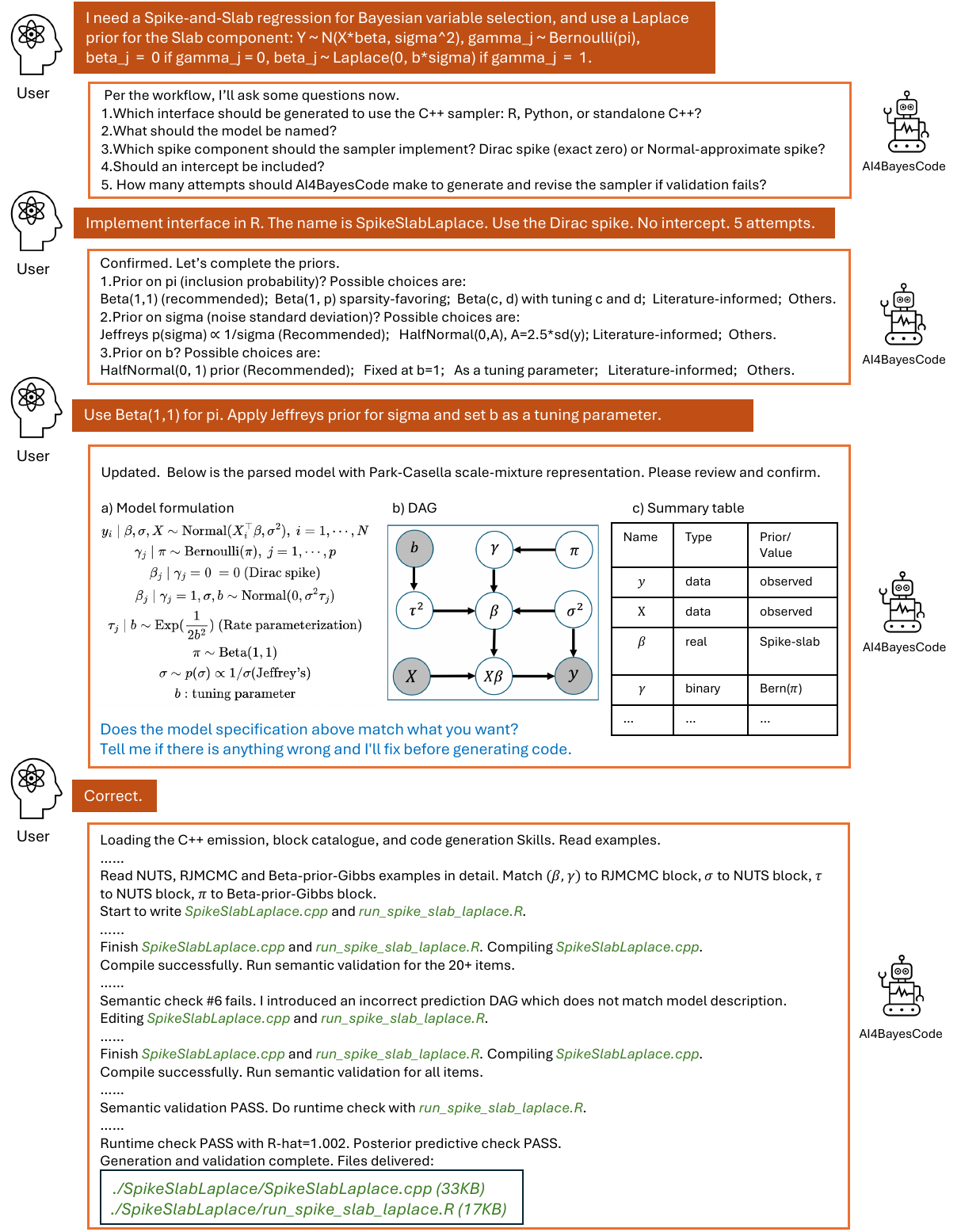}
\caption{\em An example of using AI4BayesCode to interactively generate an MCMC sampler for Dirac Spike-and-Slab regression with an AI agent. }
\label{fig:example}
\end{figure}

We present an example of using AI4BayesCode to build an MCMC sampler for the Dirac spike-and-slab model in Figure~\ref{fig:example}. {We chose this model because it highlights a model class that is not naturally handled by a standard gradient-based sampler. The Dirac spike introduces exact zeros through discrete inclusion indicators, and the Laplace slab makes marginalization difficult; AI4BayesCode handles this structure directly by assigning the coupled $(\beta,\gamma)$ update to an RJMCMC block, an update type for which many mainstream probabilistic programming languages provide limited built-in support. In Figure~\ref{fig:example}, the initial model description is intentionally informal and incomplete, reflecting how users often describe models to AI coding agents in practice; the example illustrates how AI4BayesCode can start from such realistic input and formalize the intended model before code generation. Specifically,} after the user describes the model in natural language, AI4BayesCode performs interactive pre-generation validation, including asking the user to specify the interface language, completing missing prior specifications and implementation details, and confirming the parsed model specification with the user. Once the model specification is confirmed, the system generates the sampler code and automatically performs post-generation validation. If any validation step fails, the AI agent revises the generated code based on the diagnostic feedback and retries until the sampler passes validation or the pre-specified attempt limit $M$ (default $M=5$) is reached. See the Supplementary Materials for two videos demonstrating this example through an interactive workflow with Claude Code Opus 4.7. The generated code can then be used by the user to perform statistical analyses in other interfaces. An example illustrating the use of the generated code is provided in~\ref{app:example_code}.

\section{Experiments}
\label{s:experiments}
\subsection{Setup and benchmark}

We evaluated AI4BayesCode using Claude Opus 4.7 (max effort). To assess both code generation and posterior inference across a broad range of Bayesian models, we constructed a benchmark suite of 136 models from multiple sources. This suite includes 126 models based on existing \texttt{Stan} programs 
(120 from \texttt{posteriordb} \citep{Magnusson2024}), 2 models from the \texttt{R} package \texttt{BART}, 
1 from the \texttt{R} package \texttt{BiDAG} \citep{Suter2023}, 1 from the \texttt{R} package \texttt{bayesImageS} \citep{Moores2025}, and 3 from \texttt{PyMC}. In addition, we included three generalized BART models \citep{Linero2025} for which no 
existing implementation is available. Models are grouped by category in Supplementary Materials S4. We used this benchmark suite in experiments described below.

Each benchmark model includes two components: a natural-language model description and a data-generation script. The description mimics how researchers describe a Bayesian model to an AI, and the script defines a data-generating process under which the model is correctly specified. For models drawn from existing software ecosystems, we also included the corresponding reference implementations to benchmark the code generated by AI4BayesCode.

\subsection{Experiment 1: Correct model implementation assessed by posterior agreement}
\vspace{-0.05in}
In the first experiment, we examined whether AI4BayesCode can correctly implement a Bayesian model from its natural-language description. We compared posterior inference from AI4BayesCode with that of reference implementations under repeated, correctly specified data-generating processes. For each benchmark model, we generated a single AI4BayesCode sampler using only the natural-language model description (without access to the reference implementation or the data-generating process), with an attempt limit of $M=5$. We then simulated 100 independent datasets from the corresponding data-generating process under different random seeds. For each dataset, we ran one AI4BayesCode chain and one reference chain, each with 20{,}000 burn-in iterations followed by 20{,}000 posterior draws. For each model, we report the median (over 100 replicates) of the maximum rank-normalized $\hat R$ \citep{Vehtari2021} across parameters by comparing the chains from AI4BayesCode and reference implementations, along with the median minimum bulk ESS,  the median running time, and the mean coverage of 95\% credible intervals. For the three generalized BART models, only the AI4BayesCode chain was evaluated. The running time was measured on a Mac Studio equipped with an Apple M2 Max chip. We do not report sampler generation time, as it is largely affected by network latency, the number of generation attempts, and inherent LLM variability. In practice, most models required fewer than 3 attempts and were completed within 30 minutes. 

\begin{figure}[h]
\begin{center}
\includegraphics[width=\textwidth]{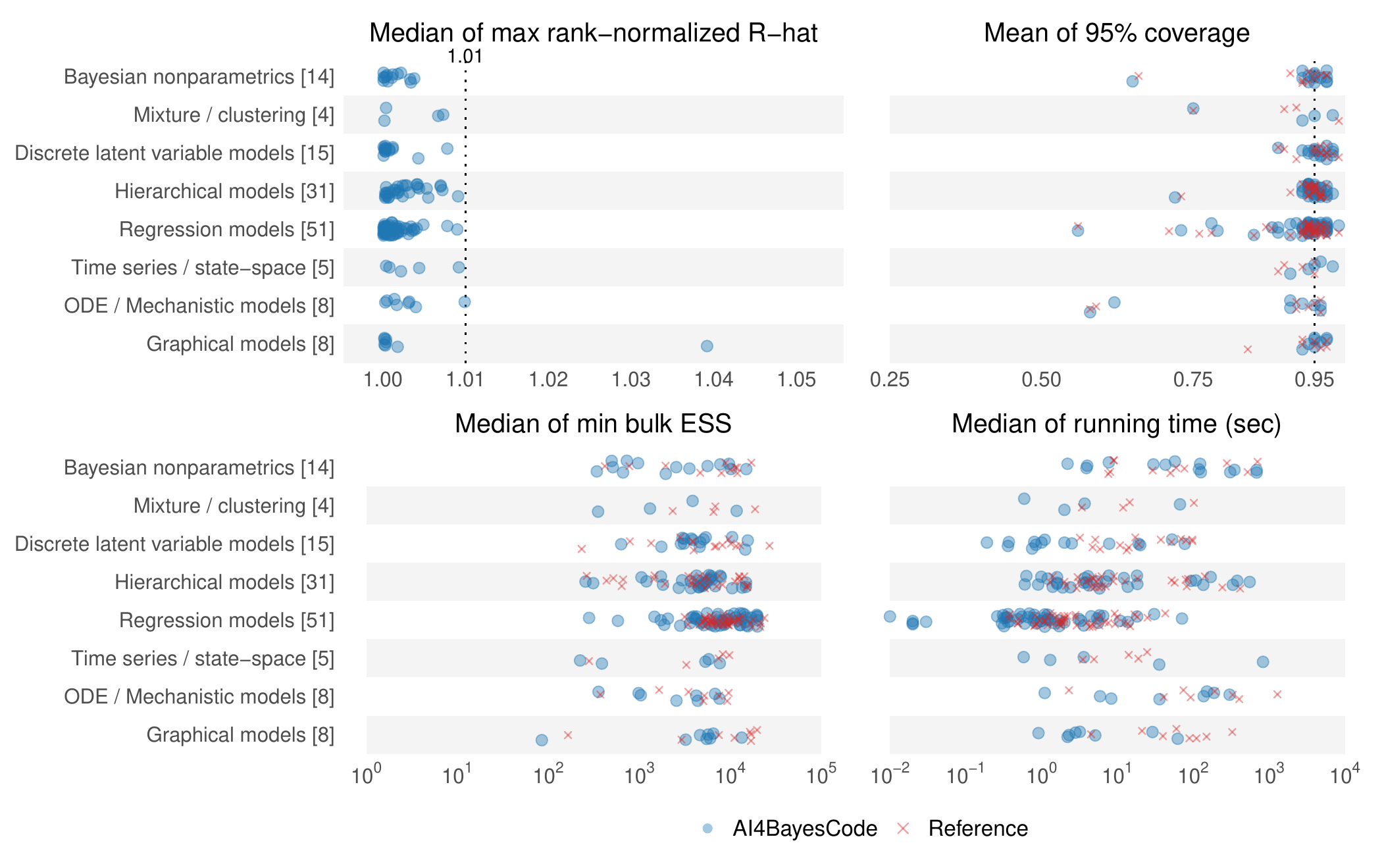}
\end{center}
\caption{\em Results of Experiment 1. Each row corresponds to a model category, and each point represents a model. Bracketed numbers indicate the number of models within each category. Red crosses represent reference samplers, where available. ESS and running time are plotted on $\log_{10}$ x-scales.} 
\label{fig:s1}
\end{figure}

Results are shown in Figure~\ref{fig:s1}, which presents model-level summaries by category. Detailed results are provided in Supplementary Materials S5. AI4BayesCode produced validated samplers that passed system validation within five attempts for all 136 benchmark models, each achieving a median maximum $\hat R \leq 1.05$, and 135 further achieved $\hat R \leq 1.01$, indicating strong posterior agreement with reference implementations. Only \texttt{water\_bif} had slightly inflated $\hat R$, since this network learning model is truly challenging. The mean coverage of 95\% credible intervals was also close to nominal for most models. AI4BayesCode yielded generally lower ESS than the reference implementations but was faster overall, particularly for models supported by specialized built-in sampling blocks. Together, these results demonstrate that AI4BayesCode can implement a wide range of Bayesian models from natural-language descriptions alone.

\subsection{Experiment 2: Variability under repeated code generation}
In the second experiment, we evaluated generation variability under repeated code generation from the same natural-language model description. To ensure computational feasibility, we selected 16 benchmark models spanning multiple categories (see~\ref{app:cateory_sim2}). For each model, we generated five independent AI4BayesCode samplers from the same natural-language model description. Each generated sampler was then evaluated on 20 independently simulated datasets from the corresponding data-generating process, with comparison to reference chains as described in Experiment 1. 

\begin{figure}[t]
\begin{center}
\includegraphics[width=\textwidth]{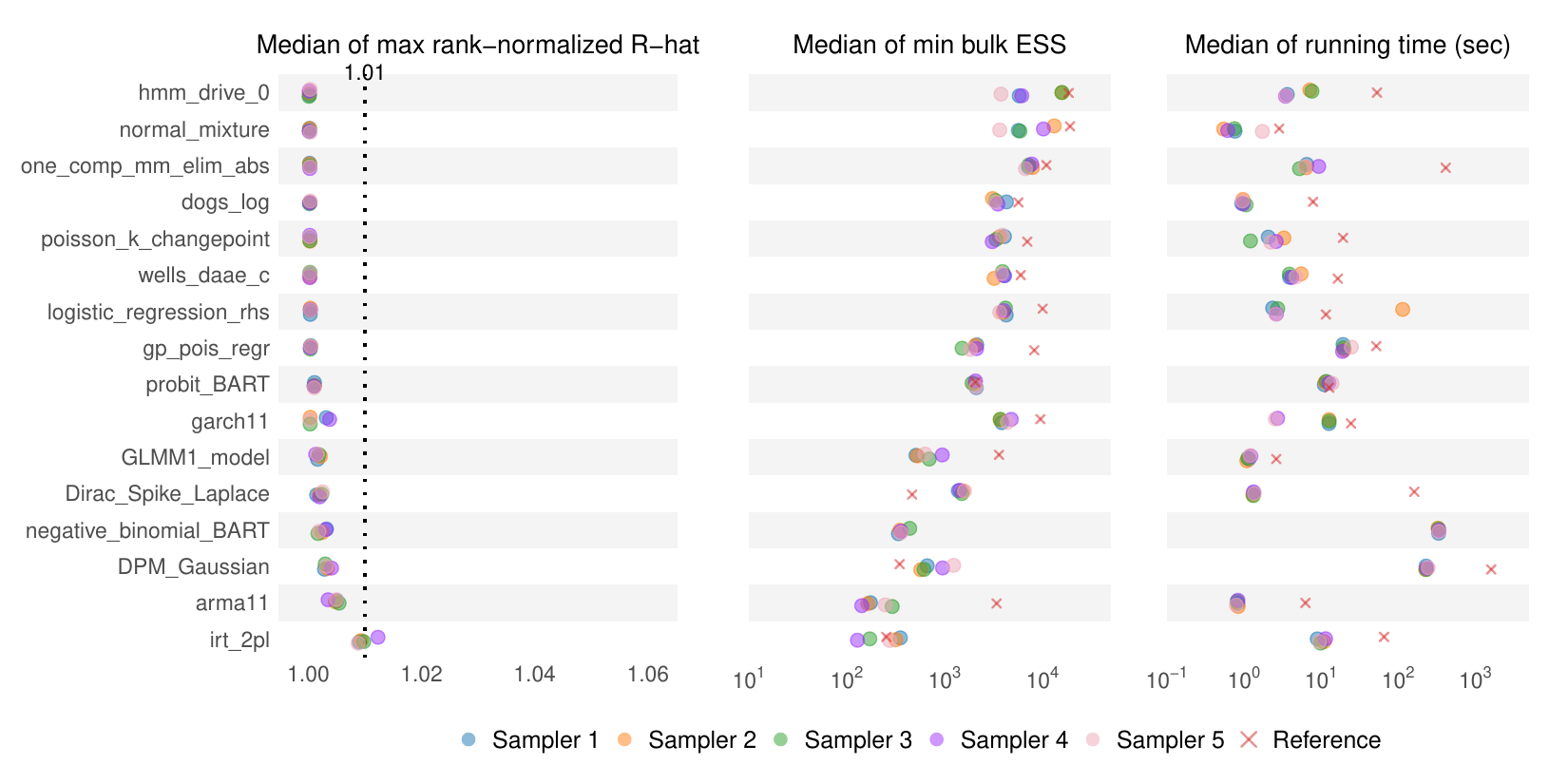}
\end{center}
\caption{\em Results of Experiment 2. 
The colored dots represent independently generated AI4BayesCode samplers, and the red crosses represent reference samplers. ESS and running time are plotted on $\log_{10}$ x-scales.}\vspace{-0.1in}
\label{fig:s2}
\end{figure}

All samplers passed the post-generation validation. For each sampler within each model, Figure \ref{fig:s2} reports the median (over 20 replicates) of the maximum rank-normalized \(\hat R\), the minimum bulk ESS, and runtime. For most models, the five independently generated samplers produced nearly identical results, with median maximum $\hat R < 1.01$. The main exception was \texttt{irt\_2pl}, for which the five generated samplers had median maximum $\hat R$ around 1.01. Same as Experiment~1, the generated samplers tended to have shorter runtimes but smaller ESS than the reference samplers. Overall, these results demonstrate strong robustness of AI4BayesCode to repeated sampler generation from the same model specification.

\subsection{Experiment 3: Validation beyond code generation alone}

\vspace{-0.05in}
\begin{table}[t]
\resizebox{\textwidth}{!}{%
\begin{tabular}{p{4.4cm}ccc c p{4.4cm}ccc}
\textbf{Model} &
\shortstack[c]{Semantic\\pass rate} &
\shortstack[c]{Runtime\\pass rate} &
\shortstack[c]{Median of \\ Maximum $\hat R$} &
&
\textbf{Model} &
\shortstack[c]{Semantic\\pass rate} &
\shortstack[c]{Runtime\\pass rate} &
\shortstack[c]{Median of \\ Maximum $\hat R$} \\
\midrule
\texttt{dogs\_log}                  & 3/5 & 3/3 & 1.00, 1.00, 1.00 &
&
\texttt{logistic\_regression\_rhs}  & 3/5 & 3/3 & 1.00, 1.00, 1.00 \\

\texttt{gp\_pois\_regr}             & 4/5 & 0/4 & --- &
&
\texttt{wells\_daae\_c\_model}             & 2/5 & 2/2 & 1.00, 1.00 \\

\texttt{poisson\_k\_changepoint}    & 4/5 & 0/4 & --- &
&
\texttt{GLMM1\_model}               & 2/5 & 1/2 & 1.00 \\

\texttt{normal\_mixture}            & 1/5 & 1/1 & 1.00 &
&
\texttt{Dirac\_Spike\_Laplace}      & 2/5 & 1/2 & 1.00 \\

\texttt{probit\_BART}               & 4/5 & 4/4 & 1.00, 1.00, 1.00, 1.00 &
&
\texttt{arma11}                     & 4/5 & 4/4 & 1.01, 1.01, 1.00, 1.00 \\

\texttt{negative\_binomial\_BART}   & 0/5 & --- & --- &
&
\texttt{garch11}                    & 3/5 & 3/3 & 1.00, 1.00, 1.00 \\

\texttt{one\_comp\_mm\_elim\_abs}   & 3/5 & 2/3 & 1.00, 1.00 &
&
\texttt{irt\_2pl}                   & 4/5 & 4/4 & 1.01, 1.01, 1.01, 1.02 \\

\texttt{hmm\_drive\_0}              & 1/5 & 0/1 & --- &
&
\texttt{DPM\_Gaussian}              & 3/5 & 0/3 & ---
\end{tabular}%
} 
\vspace{\baselineskip}
\caption{\em Results of Experiment 3. Semantic-pass rates are reported for each model across 5 generated samplers, and runtime-pass rates among the samplers that passed the semantic check. Median of maximum rank-normalized $\hat R$ is reported only from samplers that passed both checks; ``---'' indicates that no sampler passed runtime validation.}\vspace{-0.07in}
\label{tab:s3}
\end{table}

Finally, we evaluated the extent to which post-generation validation improves results beyond code generation alone. Using the same settings as in Experiment 2, we disabled semantic and runtime validations during generation, and then applied semantic and runtime validation steps separately to the generated samplers. Table~\ref{tab:s3} reports the semantic-pass rate, runtime-pass rate, and, for samplers passing both checks, the median (over 20 replicates) of the maximum rank-normalized $\hat R$. Of the 80 generated samplers from the 16 selected models, only 43 passed the semantic check and 28 passed the runtime check, indicating that systematic validation is essential to filter unqualified samplers. Among samplers that passed both validation steps, the $\hat R$ results were highly consistent with those from Experiment~2, providing further evidence of the robustness of AI4BayesCode under repeated sampler generation.



\section{Discussion}
In this paper, we introduce \textbf{AI4BayesCode}, an LLM-driven system for constructing runnable MCMC samplers from natural-language Bayesian model descriptions, supported by systematic pre-generation and post-generation validation. To evaluate the system, we develop a benchmark suite for sampler-generation assessment. Experiments show that AI4BayesCode can implement a wide range of Bayesian models from natural-language descriptions alone. Because sampler construction relies on built-in blocks, AI4BayesCode can be used to generalize code for newly proposed models by reusing and combining existing blocks; see Supplementary Materials S8. Although our experiments used Claude Code as the code-generating agent, AI4BayesCode is not tied to a particular LLM. Its codebase, AI Skills, reference examples, and validation checks can be used with other code-generating LLMs, such as OpenAI Codex (see Supplementary Materials S6 for the experiment on Codex GPT-5.5). These properties make AI4BayesCode well-suited for rapid model exploration and AI-based data-driven discovery \citep{Sun2026a, Sun2026,Duerr2026}, as its natural-language interface lowers the barrier to use and its extensible design enables rapid integration of new methods beyond existing probabilistic programming systems.

The primary design objective of AI4BayesCode is to maximize the reliability of AI-generated sampler code, even when doing so requires sacrificing sampling efficiency (e.g., reduced reliance on Gibbs sampling). For users, clear and detailed model descriptions are essential for reducing ambiguity prior to code generation. Users with sampling expertise can further improve sampler quality and potentially reduce implementation variability and runtime by specifying preferred sampling blocks or update strategies in the model description. However, silent implementation errors may still occur even when a sampler passes systematic validation. Within the harness-engineering framework \citep{Zhou2026,Pan2026,Lin2026}, AI4BayesCode supports human-in-the-loop debugging: user feedback from running validated samplers can help AI agents identify and reduce silent errors. Thus, AI4BayesCode is best suited for rapid exploration analyses; for confirmatory use, generated samplers should be further checked and, when feasible, cross-validated in established probabilistic programming systems.

Modularity and Statefulness are central design principles of AI4BayesCode. Unlike conventional probabilistic programming systems, which primarily generate samplers for standalone models, AI4BayesCode treats sampler implementations as reusable components within larger MCMC workflows. This design facilitates the reuse of code beyond its original models and allows sampling blocks developed by different contributors to be coherently integrated into larger hierarchical Bayesian models. More broadly, our work suggests a general software principle for Bayesian computation: sampler implementations should be designed not only to run standalone models, but also to serve as stateful components within larger MCMC algorithms.

Compared with probabilistic programming language systems, the main advantages of AI4BayesCode are extensibility, flexibility, and stateful composition. New model classes or sampling algorithms can be incorporated by adding built-in blocks guided by AI Skills. This is particularly useful for models that require specialized updates, such as graphical models or special Gibbs samplers. The resulting components can then be integrated into larger workflows without substantial rewriting.  This is especially useful in research settings where new models are developed by extending existing models or samplers with additional components that require complex algorithms such as NUTS. In contrast, probabilistic programming systems have two strengths: they provide reliable samplers once model specifications are correct, and their numerical computation engines are often highly optimized.

We also compared AI4BayesCode with a simpler strategy: directly asking a coding LLM to generate the sampler from the same natural-language model description, without using the AI4BayesCode workflow (see Supplementary Materials S7). The results suggest that direct prompting is already effective for simple models and for classical models with mature sampling strategies. However, for more complicated models, the direct-prompting samplers often showed worse mixing, much lower ESS (particularly with Opus 4.7), or failure to produce validated posterior computation, especially in several graphical models. In addition, these samplers generally lacked an explicit stateful sampler design, so combining them with other MCMC components would require substantial rewriting and could increase the risk of implementation errors. As coding LLMs continue to improve, direct generation of reliable samplers may become increasingly feasible; in that setting, the modular design may become less central, but stateful design and systematic validation will remain valuable.

Several challenges and future directions remain. First, weak identification remains difficult to diagnose automatically, because poor MCMC diagnostics can arise either from model non-identifiability or from implementation errors. AI4BayesCode currently flags potential weak-identification cases during validation and alerts users, but more reliable automated diagnosis remains an open problem. Second, although we have added blocks that support variational inference, the design, validation criteria, and empirical evaluation of VI-based workflows differ substantially from those for MCMC samplers; we therefore leave a systematic treatment of variational inference to future work. Finally, computational efficiency can be further improved at the kernel tier. Compared with highly optimized reference implementations such as \texttt{Stan}, AI4BayesCode remains slower in some settings. More efficient implementations of core sampling algorithms could improve sampling and runtime performance. We plan to address these limitations in future releases.




\section{Software}\label{sec:software}

AI4BayesCode itself and the samplers generated by AI4BayesCode are licensed under the GNU General Public License version 3 or later (GPL-3.0-or-later). The AI4BayesCode codebase, benchmark suite, generated sampler examples, and a platform for users to share and download user-designed blocks, are publicly available at \url{https://ai4bayescode.com/}. External numerical libraries used in the kernel tier, together with their licenses, are listed in the Supplementary Materials S9.

\section*{Acknowledgments}

The authors thank the developers and maintainers of the external numerical libraries used by AI4BayesCode.

  \bibliography{bibliography.bib}

\newpage
\appendix
\setcounter{figure}{0}
\setcounter{table}{0}
\setcounter{lstlisting}{0}

\renewcommand{\thefigure}{A\arabic{figure}}
\renewcommand{\thetable}{A\arabic{table}}
\renewcommand{\thelstlisting}{A\arabic{lstlisting}}
\renewcommand{\lstlistingname}{Code}
\renewcommand{\thesection}{Appendix \arabic{section}}

\section{Stateful composition and posterior prediction: Meta-regression example}
\label{app:meta_regression}

This appendix illustrates stateful blocks and posterior predictive sampling using a Bayesian meta-regression model with known study-specific variances. Let $y_i$ denote the observed continuous effect from study $i$, $s_i^2$ its known variance, and $x_i$ the study-level covariate vector. Consider the model \citep{Hartung2008}
\begin{equation}\label{eq:collapse_meta}
y_i \mid f_\phi,\omega,X_i,s_i^2 \sim \mathcal{N}\!\left(f_\phi(X_i),\omega^2+s_i^2\right),
\end{equation}
where $\omega^2$ is the between-study variance with half-Cauchy prior $\omega \sim \mathcal C^+(0,1)$, and $f_\phi$ is a regression function with parameters $\phi$. The function $f_\phi$ can represent a general regression model, including newly developed models without dedicated meta-regression implementations. 

\begin{figure}[h]
\begin{center}
\includegraphics[width=\textwidth]{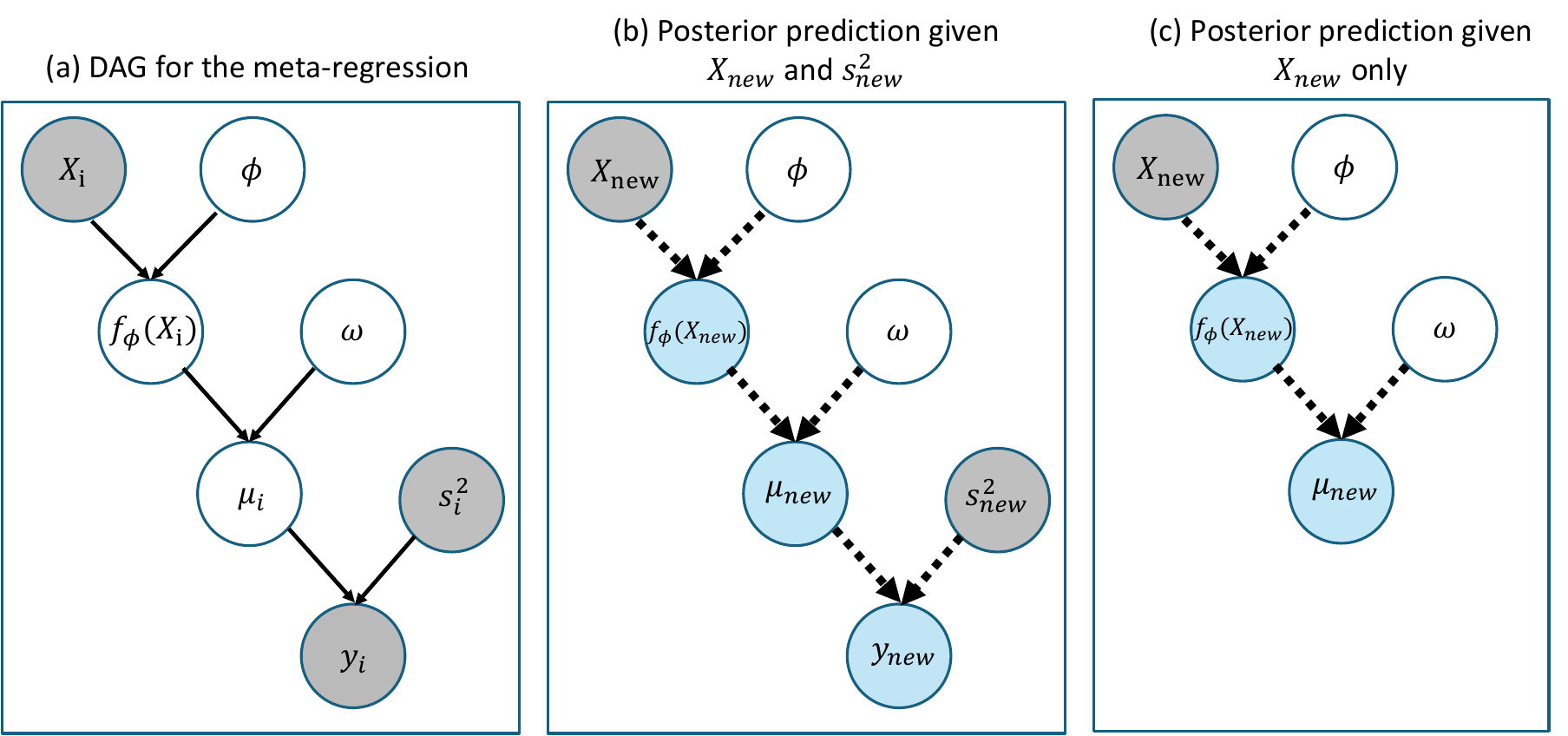}
\end{center}
\caption{\em Prediction DAGs for the meta-regression in \eqref{eq:hier_meta}. Panel (a) shows the fitted hierarchical model. Panels (b) and (c) show posterior prediction DAGs under different input settings. 
Gray nodes denote observed quantities, white nodes denote latent variables or hierarchical parameters inferred from the fitted model, and blue nodes denote posterior predictive quantities. Solid arrows indicate dependencies in the fitted model, whereas dotted arrows represent predictive propagation from upstream to downstream nodes.} 
\label{fig:meta_DAG}
\end{figure}

Model \eqref{eq:collapse_meta} can be reparameterized as 
\begin{equation}\label{eq:hier_meta}
y_i \mid \mu_i,s_i^2 \sim \mathcal{N}(\mu_i,s_i^2), \qquad
\mu_i \mid f_\phi,\omega,X_i \sim \mathcal{N}\!\left(f_\phi(X_i),\omega^2\right),
\end{equation}
where $\mu_i$ is the latent effect for study $i$. Under this formulation, $\mu_i$ follows a regression model with homoscedastic residual variance $\omega^2$. Figure~\ref{fig:meta_DAG}(a) shows the corresponding DAG, where gray nodes denote observed quantities, white nodes denote latent variables or parameters, and directed edges indicate conditional dependencies of each child node on its parent nodes.

Now consider fitting \eqref{eq:hier_meta} by MCMC. At iteration $t$, we first update $\mu_i^{(t)}$ and $\omega^{(t)}$ from their conditional posterior distributions, then sample $\phi^{(t)}$ from $\pi(\phi \mid \mu_i^{(t)},\omega^{(t)},X_i)$, using $\mu_i^{(t)}$ as the working response and $X_i$ as predictors. In a non-stateful sampler, however, the updated $\mu_i^{(t)}$ cannot be used directly without reinitializing the sampler. As a result, the $\phi$ block  does not continue from its current state but instead starts a new sampling process under the updated response $\mu_i^{(t)}$. For example, if $f_{\phi}$ is modeled using BART, its internal state is the current tree ensemble, and Bayesian backfitting updates each tree conditional on the others \citep{Chipman2010}. Reinitialization discards this ensemble, so even the first tree update is no longer conditional on the current fitted values of the other trees, but on a new initial ensemble. By contrast, a stateful block accepts the updated $\mu_i^{(t)}$ while retaining its current internal state. At iteration $t$, the $\phi$ block receives $(\mu_i^{(t)},\omega^{(t)})$ via \texttt{set\_current()}, updates to $\phi^{(t)}$ through \texttt{step()}, and returns its state through \texttt{get\_current()} for use in the next iteration.

We then consider posterior predictive sampling for the stateful sampler via the \texttt{predict\_at()} function. Each stateful sampler from AI4BayesCode internally maintains the DAG in Figure~\ref{fig:meta_DAG}(a), which distinguishes fixed quantities from latent variables and parameters, records their dependencies, and supports posterior predictions. For a new prediction task, the sampler starts from the top nodes (those without parent nodes), including any new observed inputs provided by the user. The DAG then determines which downstream nodes are predictable: a child node becomes predictable once all its parent nodes are specified, either by fitted parameters, user inputs, or previously predicted nodes. Recursively repeating this step propagates prediction through the DAG. 

This mechanism also naturally supports partial-input predictions. If only part of the required new inputs is provided, prediction proceeds only along branches whose parent nodes are fully specified. The same mechanism also supports predictions without new inputs, such as density estimation, where prediction starts from the top parameter nodes alone and propagates through the full DAG.

We illustrate this using the meta-regression model~\eqref{eq:hier_meta} in Figure~\ref{fig:meta_DAG}(b-c). In Figure~\ref{fig:meta_DAG}(b), both $X_{new}$ and $s_{new}^2$ are provided, so prediction propagates through $f_\phi(X_{new})$ to $\mu_{new}$ and then to $y_{new}$. In Figure~\ref{fig:meta_DAG}(c), only $X_{new}$ is provided. Prediction can still propagate through $f_\phi(X_{new})$ to $\mu_{new}$, but cannot proceed further to $y_{new}$, because $s_{new}^2$ is not given. If only $s_{new}^2$ is provided, no prediction is possible. Thus, under partial input, the same DAG supports prediction only along those branches whose parent nodes are fully specified.

\section{Example code}\label{app:example_code}
In this appendix section, we give a simple usage example of the generated code in Section~\ref{sec:spike_slab}. Since we selected \texttt{R} as the interface language, the generated output contains two files: \texttt{SpikeSlabLaplace.cpp} and \texttt{run\_spike\_slab\_laplace.R}. The file \texttt{SpikeSlabLaplace.cpp} contains the sampler implementation, while \texttt{run\_spike\_slab\_laplace.R} contains \texttt{R} code for runtime validation and usage examples. Due to space limits and the sampler's dependence on internal AI4BayesCode components, we do not show all implementation details. Instead, we provide an example \texttt{R} code for using the generated sampler.

\begin{lstlisting}[
%caption={\em Example \texttt{R} code for running the generated \texttt{SpikeSlabLaplace} sampler.},
label={lst:sampler-interface},
language=R,
captionpos=b,
breaklines=true,
keepspaces=true,
basicstyle={\linespread{1}\selectfont\footnotesize\ttfamily},
columns=fullflexible,
aboveskip=0.5\baselineskip,
belowskip=0.5\baselineskip
]
# Compile C++ model and source some helper functions
source("AI4BayesCode/R/AI4BayesCode_helpers.R")
AI4BayesCode_sourceCpp("./SpikeSlabLaplace/SpikeSlabLaplace.cpp",
                    AI4BayesCode_path = "AI4BayesCode")

# Simulate a toy data
N <- 100L; p <- 20L; N_pred <- 50;
X_obs  <- matrix(rnorm(N * p), N, p)                              
X_obs  <- scale(X_obs, center = TRUE, scale = TRUE)    # scale X
beta_true <- c(2.0, -1.5, 1.0, rep(0, p - 3))                      
y_obs  <- as.numeric(X_obs %*% beta_true + rnorm(N, 0, 0.5))       
y_obs  <- y_obs - mean(y_obs)                          # center y
X_test  <- matrix(rnorm(N_pred * p), N_pred, p)        # generate test-set covariates


# Monolithic chain (non-stateful use)
mono_chain = run_chain(X_obs, y = y_obs, b = 1.0, seed = 1, n_burnin = 4000, n_keep = 4000)


# Examples of stateful functions

## Initialize the SpikeSlabLaplace model. We keep the full history of posterior draws.
## If keep_history is FALSE, only the last draw is stored.
model = new(SpikeSlabLaplace, X = X_obs, y = y_obs, b = 1.0, seed = 1, keep_history = TRUE)

model$step(10)                      ## Run 10 iterations.
model$get_history()                 ## Get 10 posterior draws for all parameters.
model$step(100)                     ## Run another 100 iterations.
model$get_history()                 ## Get 110 posterior draws for all parameters.
model$get_current()                 ## Get the last (the 110th) draw for all parameters.

## Suppose y_update is the updated outcome from other blocks. 
## We use set_current() to update our sampler in a stateful way, without reinitialization.
model$set_current(list(y = y_update))
model$step(1)                           ## Run one iteration after using set_current. 
model$get_history()                     ## Get 111 posterior draws for all parameters.

## We can use predict_at() for predictions.
model$predict_at(list(X = X_test))
\end{lstlisting}

\section{Selected Models Used in Experiments 2 and 3}\label{app:cateory_sim2}
Table~\ref{tab:model_category} provides brief descriptions of the selected models used in Experiments 2 and 3, grouped by category.

\begin{table}[H]
\centering
\begingroup
\fontsize{8.0}{6.0}\selectfont
\setlength{\tabcolsep}{1.5pt}
\begin{tabular}{@{}>{\raggedright\arraybackslash}m{3.4cm}
                >{\centering\arraybackslash}m{1.3cm}
                >{\raggedright\arraybackslash}m{8.4cm}@{}}
\multicolumn{1}{c}{\textbf{Models}} &
\multicolumn{1}{c}{\textbf{Reference}} &
\multicolumn{1}{c}{\textbf{Brief description}} \\
\midrule

\multicolumn{3}{c}{\textbf{Bayesian Nonparametrics}} \\
\texttt{probit\_BART}
& \texttt{BART}
& A binary regression model with a probit link, with the regression function characterized by an additive sum of Bayesian additive regression trees. Source: \citet{Chipman2010}. \\

\shortstack[l]{\texttt{negative\_binomial}\\\texttt{\_BART}}
& ---
& A negative binomial count regression model, where the mean is a positive multiplicative function of predictors, characterized by generalized BART. Source: \citet{Linero2025}. \\

\texttt{gp\_pois\_regr}
& \texttt{Stan}
& A Poisson log-Gaussian process regression model for count data. \\

\texttt{DPM\_Gaussian}
& \texttt{PyMC}
& A Dirichlet process Gaussian mixture model for continuous data with stick-breaking weights and component-specific means and variances. \\

\midrule
\multicolumn{3}{c}{\textbf{Mixture / Clustering}} \\
\texttt{normal\_mixture}
& \texttt{Stan}
& A two-component Gaussian mixture model with unknown mixing weights and component means, and unit fixed variances. \\

\midrule
\multicolumn{3}{c}{\textbf{Discrete Latent Variable Models}} \\
\shortstack[l]{\texttt{Dirac\_Spike}\\\texttt{\_Laplace}}
& \texttt{PyMC}
& A sparse Gaussian linear regression model with a Dirac spike and Laplace slab prior on regression coefficients. \\

\shortstack[l]{\texttt{poisson\_k}\\\texttt{\_changepoint}}
& \texttt{Stan}
& A Poisson multiple changepoints model for count time series, with each segment characterized by a different rate and separated by ordered changepoints. \\

\midrule
\multicolumn{3}{c}{\textbf{Hierarchical Models}} \\
\texttt{GLMM1\_model}
& \texttt{Stan}
& A Poisson generalized linear mixed model. Source: \citet{kery2011bayesian}. \\

\texttt{irt\_2pl}
& \texttt{Stan}
& A two-parameter logistic item response theory model with person abilities and item-specific discrimination and difficulty parameters. \\

\midrule
\multicolumn{3}{c}{\textbf{Regression Models}} \\
\texttt{dogs\_log}
& \texttt{Stan}
& A binary logistic regression for dog-shock outcomes, where the probability of shock depends on each dog's cumulative previous avoidances and shocks. Source: \citet{gelman2006data}. \\

\texttt{wells\_daae\_c\_model}
& \texttt{Stan}
& A logistic regression model for the Bangladesh wells data, modeling switching behavior to covariates including distance, arsenic level, interaction, and additional household features. Source: \citet{gelman2006data}. \\

\texttt{logistic\_regression\_rhs} & \texttt{Stan} & A Bayesian logistic regression model with a regularized horseshoe prior on the regression coefficients for sparse binary-outcome modeling.\\

\midrule
\multicolumn{3}{c}{\textbf{Time Series / State-Space / Spatial Models}} \\
\texttt{arma11}
& \texttt{Stan}
& A univariate ARMA(1,1) time series model. \\

\texttt{garch11}
& \texttt{Stan}
& A univariate GARCH(1,1) time series model. \\

\midrule
\multicolumn{3}{c}{\textbf{ODE / Mechanistic Models}} \\
\shortstack[l]{\texttt{one\_comp\_mm}\\\texttt{\_elim\_abs}}
& \texttt{Stan}
& A one-compartment pharmacokinetic ODE model with first-order absorption, Michaelis-Menten elimination, and lognormal measurement error components. \\
\midrule
\multicolumn{3}{c}{\textbf{Graphical models}} \\
 \texttt{hmm\_drive\_0} & \texttt{Stan} & A $K$-state hidden Markov model for a bivariate time series, where $K$ is the number of hidden states.

\vspace{\baselineskip}
\end{tabular}
\vspace{0.4\baselineskip}
\caption{\em The 16 models used in Experiments 2 and 3, including model names, reference implementations, and brief model descriptions.}
\label{tab:model_category}
\endgroup
\end{table}

\end{document}


\def\spacingset#1{\renewcommand{\baselinestretch}%
{#1}\small\normalsize} \spacingset{1}


\if1\anon
{
  \title{\bf Supplementary Materials for \\ ``AI4BayesCode: From Natural Language Descriptions to Validated Modular Stateful Bayesian Samplers''}
\author{
Jungang Zou
\hspace{.4cm}
Alex Ziyu Jiang
\hspace{.4cm}
Qixuan Chen\thanks{
The authors gratefully acknowledge \textit{please remember to list all relevant funding sources in the version that gives all author information}
}
\\
Department of Biostatistics, Columbia University
}
  \maketitle
} \fi

\if0\anon
{
  \bigskip
  \bigskip
  \bigskip
  \begin{center}
    {\LARGE\bf Supplementary Materials for \\ ``AI4BayesCode: From Natural Language Descriptions to Validated Modular Stateful Bayesian Samplers''}
\end{center}
  \medskip
} \fi

\spacingset{1.8} 

\section{Built-in Blocks}\label{app:blocks}

Web Table~\ref{tab:appendix-blocks} summarizes the built-in blocks available in AI4BayesCode version 1.1. The example column lists one representative use case for each block, but many blocks also have additional examples not shown here.

\begingroup
\fontsize{9.0}{7.5}\selectfont
\renewcommand{\arraystretch}{0.86}
\setlength{\tabcolsep}{3pt}

\setlength{\LTcapwidth}{\textwidth}
\captionsetup[longtable]{
    justification=raggedright,
    singlelinecheck=false
}

\begin{longtable}{@{}>{\raggedright\arraybackslash}p{4.7cm}
                     >{\raggedright\arraybackslash}p{\dimexpr\textwidth-4.7cm-3.5cm-4\tabcolsep\relax}
                     >{\raggedright\arraybackslash}p{3.5cm}@{}}

\caption{Built-in blocks available in AI4BayesCode. The last column gives one representative example for each block.}
\label{tab:appendix-blocks} \\

\toprule
\textbf{Block} & \textbf{Description} & \textbf{Example} \\
\midrule
\endfirsthead

\multicolumn{3}{@{}l@{}}{\small\itshape (Table~\thetable\ continued from previous page)} \\
\toprule
\textbf{Block} & \textbf{Description} & \textbf{Example} \\
\midrule
\endhead

\midrule
\multicolumn{3}{r@{}}{\small\itshape continued on next page} \\
\endfoot

\bottomrule
\endlastfoot

\multicolumn{3}{c}{\textbf{Generic transition kernels}} \\
\texttt{nuts} & NUTS on a single named parameter \citep{Hoffman2014} & \texttt{GaussianLocationScale} \\
\texttt{joint\_nuts} & Multi-parameter joint NUTS \citep{Hoffman2014} & \texttt{BSplineRegression} \\
\texttt{univariate\_slice\_sampling} & 1-D slice sampling \citep{neal2003slice} & \texttt{GPTimeSeries} \\
\texttt{elliptical\_slice\_sampling} & Elliptical slice sampling \citep{Murray2010} & \texttt{GPClassification} \\
\texttt{cluster\_atom}
& Slice-sampling update for mixture-model atoms with nonconjugate priors \citep{Ishwaran2001,neal2003slice} & \texttt{DPGammaMixture} \\
\midrule

\multicolumn{3}{c}{\textbf{Discrete-latent Gibbs}} \\
\texttt{binary\_gibbs} & Closed-form Bernoulli vector & --- \\
\texttt{categorical\_gibbs} & Closed-form $K$-way categorical & \texttt{FiniteGaussianMixture} \\
\texttt{lda\_collapsed\_gibbs} & Collapsed Gibbs for LDA \citep{Griffiths2004} & \texttt{LdaCollapsedGibbs} \\
\texttt{stick\_breaking} & Truncated stick-breaking for DP / PY / custom BNP \citep{Ishwaran2001} & \texttt{DPGaussianMixture} \\
\midrule

\multicolumn{3}{c}{\textbf{Continuous-conjugate Gibbs}} \\
\texttt{beta\_gibbs} & Conjugate Beta scalar & \texttt{SpikeSlabRJMCMC} \\
\texttt{dirichlet\_gibbs} & Conjugate Dirichlet simplex & \texttt{FiniteGaussianMixture} \\
\texttt{gamma\_gibbs} & Conjugate Gamma scalar & --- \\
\texttt{inv\_gamma\_gibbs} & Conjugate inverse-Gamma scalar & --- \\
\texttt{normal\_gamma\_cluster\_gibbs} & Diagonal NG cluster sampler & \texttt{FiniteGaussianMixture} \\
\texttt{niw\_cluster\_gibbs} & Full-covariance NIW cluster sampler & \texttt{HDPGaussianMixture} \\
\texttt{celerite\_gp} & 1-D celerite Gaussian process \citep{ForemanMackey2017} & \texttt{GPTimeSeries} \\
\midrule

\multicolumn{3}{c}{\textbf{Data-augmentation Gibbs}} \\
\texttt{pg\_logistic} & Pólya-Gamma logistic augmentation \citep{Polson2013} & \texttt{LogisticRegression} \\
\texttt{probit\_aug} & Albert--Chib probit data augmentation \citep{Albert1993} & \texttt{ProbitRegression} \\
\texttt{poisson\_multinomial\_aug} & Poisson augmentation for multinomial models & \texttt{GBartMultinomial} \\
\midrule

\multicolumn{3}{c}{\textbf{Survival-specific blocks}} \\
\texttt{piecewise\_exponential\_gibbs}
& Gibbs update for piecewise-exponential baseline hazards \citep{Ibrahim2001}
& \texttt{PehSurvival} \\

\begin{tabular}[t]{@{}l@{}}
\texttt{interval\_censored\_}\\
\texttt{survival\_augmentation}
\end{tabular}
& Latent-time augmentation for interval-censored PEH survival data \citep{Ibrahim2001}
& --- \\

\texttt{frailty\_gamma\_gibbs}
& Gibbs update for shared Gamma frailties \citep{Ibrahim2001}
& \texttt{PehSharedFrailty} \\
\midrule

\multicolumn{3}{c}{\textbf{Trans-dimensional MH}} \\
\texttt{rjmcmc} & Reversible-jump MCMC \citep{Green1995} & \texttt{SpikeSlabRJMCMC} \\
\texttt{split\_merge} & Jain--Neal split-merge MH \citep{jain2004split} & --- \\
\midrule

\multicolumn{3}{c}{\textbf{Specialized algorithms for Bayesian graphical models}} \\
\texttt{hmm}
& FFBS for finite-state HMM \citep{FruhwirthSchnatter2006}
& \texttt{HMMGaussian2State} \\

\texttt{ising\_cluster}
& Swendsen--Wang cluster sampler for Ising / Potts MRFs \citep{Swendsen1987}
& \texttt{IsingPrior} \\

\texttt{gmrf\_precision}
& Sparse-Cholesky direct sampler for Gaussian MRFs \citep{Rue2001}
& \texttt{ICARSpatialGMRF} \\

\texttt{gmrf\_whitened\_ess}
& GMRF prior with non-Gaussian likelihood via elliptical slice sampling \citep{Murray2010,Rue2001}
& --- \\

\texttt{gmrf\_gaussian\_joint} & Joint sampler for Gaussian GMRFs with unknown smoothing precision \citep{KnorrHeld2002} & --- \\

\texttt{order\_mcmc}
& Order MCMC for Bayesian-network structure learning \citep{Heckerman1995,Friedman2003}
& \texttt{OrderMCMCBN} \\
\midrule

\multicolumn{3}{c}{\textbf{Tree-ensemble priors}} \\
\texttt{bart}
& Gaussian BART \citep{Chipman2010}
& \texttt{BartNoise} \\

\texttt{genbart}
& Generalized BART via RJMCMC \citep{Linero2025}
& \texttt{GBartPoisson} \\

\texttt{softbart}
& Soft BART \citep{Linero2018}
& \texttt{SoftBartNoise} \\
\midrule

\multicolumn{3}{c}{\textbf{Variational inference}} \\
\texttt{mean\_field\_gaussian\_vi}
& Mean-field Gaussian ADVI \citep{Kucukelbir2017}
& --- \\

\texttt{full\_rank\_gaussian\_vi}
& Full-rank Gaussian ADVI \citep{Kucukelbir2017}
& --- \\

\texttt{mean\_field\_categorical\_vi}
& Mean-field VI for discrete latents
& \texttt{CategoricalIsingChainVI} \\

\texttt{structured\_categorical\_vi}
& Clique-structured VI \citep{Saul1995}
& \texttt{StructuredPottsVI} \\
\midrule

\multicolumn{3}{c}{\textbf{Tools and helpers}} \\
\texttt{composite\_block} & Container combining child blocks & \texttt{GaussianLocationScale} \\
\texttt{block\_sampler} & Abstract base class & --- \\
\texttt{vi\_block} & Abstract base class for VI & --- \\
\texttt{vi\_optimizer} & RAABBVI \citep{Welandawe2024} & --- \\
\texttt{constraints} & Parameter transforms & \texttt{GaussianLocationScale} \\
\texttt{shared\_data} & Inter-block data store & --- \\
\texttt{ode\_rk45} & Adaptive RK45 ODE integrator & \texttt{ODE\_SIR} \\
\texttt{ode\_linear} & Exact linear ODE solver & --- \\
\texttt{rjmcmc\_transforms} & Trans-dimensional move helpers & \texttt{SpikeSlabRJMCMC} \\
\texttt{celerite\_marginal\_likelihood} & Celerite GP marginal likelihood & \texttt{GPTimeSeries} \\
\texttt{bnp\_utils} & BNP helpers, including stick-breaking and weights & \texttt{FiniteGaussianMixture} \\
\texttt{bde\_scorer} & BDe / BDeu family scorer for BN structure learning & \texttt{OrderMCMCBN} \\
\texttt{score\_cache} & Cached local scores for BN structure learning & \texttt{OrderMCMCBN} \\
\texttt{bge\_scorer} & BGe scoring for continuous BN structure learning \citep{Geiger2002} & --- \\
\texttt{i\_scorer} & Decomposable local scoring for BN structure learning & --- \\
\texttt{partition\_state} & Partition representation for partition MCMC \citep{Kuipers2017} & \texttt{OrderMCMCBN} \\
\texttt{rjmcmc\_custom\_bijection} & Custom RJMCMC bijection with automatic Jacobian & \texttt{SpikeSlabSinhBijection} \\
\texttt{autodiff\_wrap} & Autodiff wrapper for validator & --- \\
\texttt{pybind\_casters} & \texttt{Python} type casters & --- \\
\texttt{rcpp\_wrap} & \texttt{R}/\texttt{Rcpp} wrappers & --- \\
\texttt{backend\_neutral} & Shared utilities for \texttt{R} and \texttt{Python} interfaces & --- \\
\texttt{kernel\_control\_mixin} & Kernel state control utilities & --- \\
\texttt{rcpp\_predict\_guard} & Preserves R matrix shape through \texttt{predict\_at()} & --- \\
\texttt{types} & Type aliases & --- \\

\end{longtable}
\endgroup

\section{Semantic Validation Checklist}
\label{app:l2_registry}
The semantic check in post-generation validation step verifies a list of more than twenty items of the generated code. Web Table~\ref{tab:l2_registry} lists the full checklist with a simple description. 

We briefly explain several checks in Web Table~\ref{tab:l2_registry}. Check~12 concerns gradient verification. Ideally, the gradient of the log posterior density would be computed directly by automatic differentiation tools, such as \texttt{autodiff} \citep{Leal2024}, rather than written by AI, which may potentially introduce errors. In practice, however, calling a general automatic differentiation library at every MCMC iteration is often too expensive. AI4BayesCode therefore allows AI agent to write the gradient explicitly, and the validator immediately compares this AI-written gradient against the corresponding gradient computed by \texttt{autodiff} at a set of test points during validation. In this way, \texttt{autodiff} is used for validation, but not inside every sampling step.

Checks~16 and 17 concern Gibbs updates. In AI4BayesCode, Gibbs updates are assigned the lowest priority and are used only when necessary. Therefore, whenever a Gibbs update is used, Check~16 requires the generated code to document in code comments why Gibbs is appropriate in that case, and Check~17 restricts AI-written Gibbs samplers to approved special cases only.

Checks~18 and 20 concern the adaptation settings in NUTS. Dense-metric NUTS is computationally more expensive than diagonal-metric NUTS and therefore should be used only when it is truly needed, with enough pilot warmup to estimate the metric reliably. Check~20 requires warmup to occur only at initialization. In a stateful NUTS block, the same block is called repeatedly across outer MCMC iterations while carrying its current state forward. Re-entering warmup at each call would repeatedly re-adapt the block under changing conditioning values, which violates the stateful design. Check~23 concerns \texttt{readapt\_NUTS}, an on-demand re-adaptation routine: when a downstream sampler shifts the conditional posterior, the user may re-tune the metric without advancing chain state. The check verifies that this routine leaves state untouched and uses a separate RNG stream.

Checks~21--22 concern variational-inference (VI) blocks, which AI4BayesCode supports as an alternative to MCMC. Check~21 ensures each VI block conforms to the abstract \texttt{vi\_block} interface; Check~22 requires the optimizer to be the recommended RAABBVI variant \citep{Welandawe2024}.

\begingroup
\fontsize{8.0}{8.0}\selectfont
\renewcommand{\arraystretch}{0.92}
\setlength{\tabcolsep}{3pt}

\setlength{\LTcapwidth}{\textwidth}
\captionsetup[longtable]{
    justification=raggedright,
    singlelinecheck=false
}

\begin{longtable}{@{}
    >{\raggedright\arraybackslash}p{0.035\textwidth}
    >{\raggedright\arraybackslash}p{0.33\textwidth}
    >{\raggedright\arraybackslash}p{\dimexpr\textwidth-0.035\textwidth-0.33\textwidth-4\tabcolsep\relax}
    @{}}

\caption{Checklist of semantic validation used in AI4BayesCode.}
\label{tab:l2_registry} \\

\toprule
\textbf{\#} & \textbf{Name} & \textbf{Simple description} \\
\midrule
\endfirsthead

\multicolumn{3}{@{}l@{}}{\small\itshape
(Table~\thetable\ continued from previous page)} \\
\toprule
\textbf{\#} & \textbf{Name} & \textbf{Simple description} \\
\midrule
\endhead

\midrule
\multicolumn{3}{r@{}}{\small\itshape continued on next page} \\
\endfoot

\bottomrule
\endlastfoot

1 & Distribution parameterization & Checks whether each distribution is written with the correct parameterization. \\

2 & Sequential update & Checks whether parameters are updated in the right order. \\

3 & Dead parameters & Checks whether every parameter is actually updated. \\

4 & Regression intercept / offset & Checks whether a user-specified intercept or offset term has been omitted. \\

5 & Jacobian handling & Checks whether constrained parameters are transformed correctly. \\

6 & DAG consistency & Checks whether the DAG in the code matches the model structure. \\

7 & Dependency declaration & Checks whether each block declares the quantities it depends on. \\

8 & Rcpp API correctness & Checks whether the Rcpp interface is written correctly. \\

9 & Numerical stability & Checks for numerical problems such as overflow, invalid logs, or division by zero. \\

10 & State mutation in \texttt{predict\_at()} & Checks whether posterior prediction accidentally changes values of parameters or latent variables. \\

11 & Joint NUTS block extra audit & Adds extra checks when several parameters are updated together in one NUTS block. \\

12 & Gradient verification via \texttt{autodiff} & Checks whether the AI-written gradient matches an autodiff reference. \\

13 & RNG separation & Checks whether MCMC sampling and posterior prediction use separate random-number streams. \\

14 & Bijection-consistency AD check & Checks whether an RJMCMC bijection is implemented consistently. \\

15 & Library parity test & Checks whether a built-in block matches the intended target update. \\

16 & Inline Gibbs-exception justification & Checks whether any Gibbs exception is explicitly explained. \\

17 & No AI-written Gibbs outside whitelist & Checks that AI-written Gibbs samplers only appear in approved cases. \\

18 & Dense metric justification + pilot scaling & Checks whether dense-metric NUTS is needed and uses enough pilot warmup. \\

19 & Vectorized gradient computation & Checks whether matrix-based gradients are computed in an efficient vectorized form. \\

20 & No per-step warmup in NUTS blocks & Checks that the sampler does not keep re-entering warmup during MCMC updates. \\

21 & VI block contract conformance & Checks whether each VI block follows the framework's design: subclassing the abstract \texttt{vi\_block}, declaring its engine kind, and writing posterior samples back to history with the right shape. \\

22 & VI optimizer is RAABBVI & Checks whether the VI optimizer is the standard RAABBVI variant (averaged-Adam + iterate averaging + R-hat-based convergence + symmetrized-KL termination). \\

23 & \texttt{readapt\_NUTS()} state-preservation & Checks that the on-demand metric re-adaptation call leaves chain state unchanged and uses a separate RNG stream from the main sampler. \\

24 & Joint-NUTS pathology pre-flight & Checks a joint NUTS block for three known pathologies: a centered funnel, a constraint not matching the declared support, and a sub-parameter never read in the log-density (silently sampled from its prior). \\

25 & Trans-dimensional / Dirac-spike RJMCMC & Checks that Dirac point-mass spike or trans-dimensional models are routed to \texttt{rjmcmc\_block}, avoiding reducible Gibbs updates and silent slab-only NUTS marginalization. \\

26 & Kernel-control conformance & Checks whether the wrapper exposes the \texttt{freeze} / \texttt{unfreeze} / \texttt{get\_frozen} interface, refuses unsupported blocks, and holds a frozen parameter fixed. \\

\end{longtable}
\endgroup

\section{Runtime Validation Checklist}
\label{app:l3_registry}
Web Table~\ref{tab:l3_registry} lists the full runtime validation checklist.

\begingroup
\fontsize{9.0}{10.0}\selectfont
\renewcommand{\arraystretch}{1.22}
\setlength{\tabcolsep}{3pt}

\setlength{\LTcapwidth}{\textwidth}
\captionsetup[longtable]{
    justification=raggedright,
    singlelinecheck=false
}

\begin{longtable}{@{}
    >{\raggedright\arraybackslash}p{1.2cm}
    >{\raggedright\arraybackslash}p{3cm}
    >{\raggedright\arraybackslash}p{\dimexpr\textwidth-0.6cm-3.6cm-4\tabcolsep\relax}
    @{}}

\caption{Checklist of runtime validation used in AI4BayesCode. Rank-normalized $\widehat R$ and posterior predictive summary checks determine validation pass/fail status, whereas ESS (except for the failure rule in R2.2) and PSIS-LOO Pareto-$\hat{k}$ are used as diagnostic summaries for warning.}
\label{tab:l3_registry} \\

\toprule
\textbf{\#} & \textbf{Name} & \textbf{Simple description} \\
\midrule
\endfirsthead

\multicolumn{3}{@{}l@{}}{\small\itshape
(Table~\thetable\ continued from previous page)} \\
\toprule
\textbf{\#} & \textbf{Name} & \textbf{Simple description} \\
\midrule
\endhead

\midrule
\multicolumn{3}{r@{}}{\small\itshape continued on next page} \\
\endfoot

\bottomrule
\endlastfoot

R1  & Smoke test & Runs $10$ MCMC steps and checks that the sampler's stateful functions work. \\

R2.1  & Two-chain $\widehat R$ convergence &
Runs two parallel chains with 4{,}000 burn-in iterations and 4{,}000 posterior draws, extended to 20{,}000 burn-in iterations and 20{,}000 posterior draws if the check fails. Validation requires rank-normalized $\widehat R < 1.05$ for every parameter. \\

R2.2 & Effective-sample-size diagnostic & Reports the ESS ratio  $r = \min(\mathrm{ESS}_{\mathrm{bulk}}, \mathrm{ESS}_{\mathrm{tail}}) / n_{\mathrm{draw}}$ from the two chains; warns when $0.005 \leq r < 0.01$, and when $r < 0.005$ re-runs with 20{,}000 burn-in iterations and 20{,}000 draws, failing only if $r$ is still below $0.005$. \\

R3.1 & Posterior predictive check & Computes summary statistics for the observed data and the corresponding posterior predictive draws. For continuous variables, the summaries include the mean, standard deviation, minimum, maximum, $q_{25}$, and $q_{75}$; for binary variables, the mean; and for count variables, the mean, standard deviation, and maximum. The validation decision is based on the mean, standard deviation, $q_{25}$, and $q_{75}$, when applicable, while the minimum and maximum are reported for reference but are not used to determine pass or failure because they can be highly variable even under a well-calibrated model. The check fails if at least two evaluated summaries fall outside their corresponding 90\% posterior predictive intervals and issues a warning if exactly one does. For binary variables, the mean is the sole evaluated summary and therefore determines the result directly. When a joint NUTS block is used, the posterior predictive interval is tightened to 80\%. \\

R3.2 & PSIS-LOO Pareto-$\hat{k}$ diagnostic & Computes PSIS-LOO \citep{Vehtari2024} and summarizes the distribution of Pareto-$\hat{k}$ values across observations as a diagnostic of leave-one-out cross-validation reliability. The diagnostic does not affect validation pass/fail status, but issues a warning if fewer than 50\% of observations have $\hat{k} < 0.5$ or if at least 10\% have $\hat{k} \geq 0.7$. \\

\end{longtable}
\endgroup

\section{Model Categories}\label{app:cateory}
Table~\ref{tab:model_category} summarizes the benchmark models used in our experiments, grouped into broad model categories. Each model is assigned to a single category for reporting purposes. Among these 136 models, \texttt{Dirac\_Spike\_Laplace}, \texttt{Dirac\_Spike\_normal}, and \texttt{DPM\_Gaussian} use \texttt{PyMC} as the reference implementation, while \texttt{continuous\_BART} and \texttt{probit\_BART} use the \texttt{BART} \texttt{R} package as the reference implementation. For graphical models, we use \texttt{R} packages \texttt{bayesImageS} and \texttt{BiDAG} for \texttt{hex\_ising\_seg\_0} and \texttt{water\_bif}, respectively.

For the three generalized BART models, \texttt{logistic\_BART}, \texttt{poisson\_BART}, and \texttt{negative\_binomial\_BART}, we did not identify existing packages that match the target model specifications. Although the \texttt{BART} R package implements logistic BART, its data-augmentation scheme uses a fixed leaf-prior scale, whereas our generalized BART specification follows the half-Cauchy hyperprior in \citet{Linero2025}. We therefore do not include a direct reference comparison for this model.

\begingroup
\fontsize{9.0}{9.8}\selectfont
\setlength{\tabcolsep}{4pt}
\renewcommand{\arraystretch}{0.90}

\begin{longtable}{@{}>{\raggedright\arraybackslash}p{3.1cm}
                  >{\raggedright\arraybackslash}p{12.3cm}@{}}

\textbf{Category} & \textbf{Models} \\
\midrule
\endfirsthead

\multicolumn{2}{l}{\emph{Continued from previous page}} \\
\textbf{Category} & \textbf{Models} \\
\midrule
\endhead

Bayesian nonparametrics &
\texttt{continuous\_BART}, \texttt{logistic\_BART}, \texttt{poisson\_BART}, \texttt{negative\_binomial\_BART}, \texttt{probit\_BART}, \texttt{DPM\_Gaussian}, \texttt{GP\_normal}, \texttt{gp\_regr}, \texttt{gp\_pois\_regr}, \texttt{hierarchical\_gp}, \texttt{kronecker\_gp}, \texttt{accel\_gp}, \texttt{nn\_rbm1bJ10}, \texttt{nn\_rbm1bJ100} \\
\midrule
Mixture / clustering &
\texttt{normal\_mixture}, \texttt{normal\_mixture\_k}, \texttt{low\_dim\_gauss\_mix}, \texttt{low\_dim\_gauss\_mix\_collapse} \\
\midrule
Discrete latent variable models &
\texttt{Dirac\_Spike\_Laplace}, \texttt{Dirac\_Spike\_normal}, \texttt{ldaK2}, \texttt{ldaK5}, \texttt{poisson\_changepoint}, \texttt{poisson\_k\_changepoint}, \texttt{iohmm\_reg}, \texttt{M0\_model}, \texttt{Mb\_model}, \texttt{Mh\_model}, \texttt{Mt\_model}, \texttt{Mtbh\_model}, \texttt{Mth\_model}, \texttt{multi\_occupancy}, \texttt{Survey\_model} \\
\midrule
Time series / state-space &
\texttt{arK}, \texttt{arma11}, \texttt{garch11}, \texttt{prophet}, \texttt{state\_space\_stochastic\_level\_stochastic\_seasonal} \\
\midrule
ODE / Mechanistic models &
\texttt{sir}, \texttt{covid19imperial\_v2}, \texttt{covid19imperial\_v3}, \texttt{lotka\_volterra}, \texttt{one\_comp\_mm\_elim\_abs}, \texttt{soil\_incubation}, \texttt{dugongs\_model}, \texttt{losscurve\_sislob} \\
\midrule
Hierarchical models & \texttt{GLMM\_Poisson\_model}, \texttt{GLMM1\_model}, \texttt{bones\_model}, {\texttt{dirichlet\_meanprec\_hier}}, \texttt{eight\_schools\_centered}, \texttt{eight\_schools\_noncentered}, \texttt{election88\_full}, \texttt{pilots}, \texttt{rats\_model}, \texttt{surgical\_model}, \texttt{radon\_county}, \texttt{radon\_county\_intercept}, \texttt{radon\_hierarchical\_intercept\_centered}, \texttt{radon\_hierarchical\_intercept\_noncentered}, \texttt{radon\_partially\_pooled\_centered}, \texttt{radon\_partially\_pooled\_noncentered}, \texttt{radon\_variable\_intercept\_centered}, \texttt{radon\_variable\_intercept\_noncentered}, \texttt{radon\_variable\_intercept\_slope\_centered}, \texttt{radon\_variable\_intercept\_slope\_noncentered}, \texttt{radon\_variable\_slope\_centered}, \texttt{radon\_variable\_slope\_noncentered}, \texttt{seeds\_centered\_model}, \texttt{seeds\_model}, \texttt{seeds\_stanified\_model}, \texttt{pl2\_latent\_reg\_irt}, \texttt{gpcm\_latent\_reg\_irt}, \texttt{grsm\_latent\_reg\_irt}, \texttt{hier\_2pl}, \texttt{irt\_2pl}, \texttt{lsat\_model} \\
\midrule
Regression models &
\texttt{blr}, \texttt{diamonds}, \texttt{earn\_height}, \texttt{kilpisjarvi}, \texttt{radon\_pooled}, \texttt{kidscore\_interaction}, \texttt{kidscore\_interaction\_c}, \texttt{kidscore\_interaction\_c2}, \texttt{kidscore\_interaction\_z}, \texttt{kidscore\_mom\_work}, \texttt{kidscore\_momhs}, \texttt{kidscore\_momhsiq}, \texttt{kidscore\_momiq}, \texttt{mesquite}, \texttt{logmesquite}, \texttt{logmesquite\_logva}, \texttt{logmesquite\_logvas}, \texttt{logmesquite\_logvash}, \texttt{logmesquite\_logvolume}, \texttt{log10earn\_height}, \texttt{logearn\_height}, \texttt{logearn\_height\_male}, \texttt{logearn\_interaction}, \texttt{logearn\_interaction\_z}, \texttt{logearn\_logheight\_male}, \texttt{nes}, \texttt{sesame\_one\_pred\_a}, \texttt{GLM\_Binomial\_model}, \texttt{dogs}, \texttt{dogs\_hierarchical}, \texttt{dogs\_log}, \texttt{dogs\_nonhierarchical}, \texttt{nes\_logit\_model}, \texttt{wells\_daae\_c\_model}, \texttt{wells\_dae\_c\_model}, \texttt{wells\_dae\_inter\_model}, \texttt{wells\_dae\_model}, \texttt{wells\_dist}, \texttt{wells\_dist100\_model}, \texttt{wells\_dist100ars\_model}, \texttt{wells\_interaction\_c\_model}, \texttt{wells\_interaction\_model}, \texttt{Rate\_1\_model}, \texttt{Rate\_2\_model}, \texttt{Rate\_3\_model}, \texttt{Rate\_4\_model}, \texttt{Rate\_5\_model}, \texttt{GLM\_Poisson\_model}, \texttt{two\_normal\_spike\_slab\_model}, \texttt{logistic\_regression\_rhs}, \texttt{accel\_splines} \\
\midrule
Graphical models & \texttt{bym2\_offset\_only}, \texttt{hmm\_drive\_0}, \texttt{hmm\_drive\_1}, \texttt{hmm\_example}, \texttt{hmm\_gaussian}, {\texttt{hex\_ising\_seg\_0}}, {\texttt{pois\_hmm\_k3\_0}}, {\texttt{water\_bif}}\\
\midrule

\caption{\em Benchmark models grouped into categories. Each model has a natural-language model description, a data-generation script, and a reference model implementation (except \texttt{logistic\_BART}, \texttt{poisson\_BART}, \texttt{negative\_binomial\_BART} from the generalized BART family). }
\label{tab:model_category}\\

\end{longtable}

\endgroup

\section{Results of Experiment 1}\label{supp:exp1}
The detailed results of each model in Experiment 1 are listed in Web Table~\ref{tab:sim1_results_all}.

\begingroup
\fontsize{9.0}{9.6}\selectfont
\renewcommand{\arraystretch}{0.86}
\setlength{\tabcolsep}{3pt}

\setlength{\LTcapwidth}{\textwidth}
\captionsetup[longtable]{
    justification=raggedright,
    singlelinecheck=false
}

\begin{longtable}{@{}
    >{\raggedright\arraybackslash}p{6.7cm}
    >{\centering\arraybackslash}p{1.35cm}
    >{\centering\arraybackslash}p{1.05cm}
    >{\centering\arraybackslash}p{1.05cm}
    >{\centering\arraybackslash}p{1.15cm}
    >{\centering\arraybackslash}p{1.15cm}
    >{\centering\arraybackslash}p{1.15cm}
    >{\centering\arraybackslash}p{1.15cm}
    @{}}

\caption{Per-model results for Experiment 1, grouped by category. For each benchmark model, we summarize 100 replicate datasets by reporting the median of the maximum rank-normalized $\hat{R}$ over all parameters, the mean coverage of 95\% credible intervals across parameters, the median of the minimum bulk effective sample size, and the median running time. Generalized BART models have no reference implementation and report ``---'' in the reference columns.}
\label{tab:sim1_results_all} \\

\toprule
\multirow{2}{*}{\textbf{Model}} &
\multirow{2}{*}{\shortstack[c]{\textbf{Median}\\\textbf{max} $\boldsymbol{\hat{R}}$}} &
\multicolumn{2}{c}{\textbf{Mean coverage}} &
\multicolumn{2}{c}{\textbf{Min ESS}} &
\multicolumn{2}{c}{\textbf{Runtime (s)}} \\
\cmidrule(lr){3-4}
\cmidrule(lr){5-6}
\cmidrule(lr){7-8}
& & AI & REF & AI & REF & AI & REF \\
\midrule
\endfirsthead

\multicolumn{8}{@{}l@{}}{\small\itshape
(Table~\thetable\ continued from previous page)} \\
\toprule
\multirow{2}{*}{\textbf{Model}} &
\multirow{2}{*}{\shortstack[c]{\textbf{Median}\\\textbf{max} $\boldsymbol{\hat{R}}$}} &
\multicolumn{2}{c}{\textbf{Mean coverage}} &
\multicolumn{2}{c}{\textbf{Min ESS}} &
\multicolumn{2}{c}{\textbf{Runtime (s)}} \\
\cmidrule(lr){3-4}
\cmidrule(lr){5-6}
\cmidrule(lr){7-8}
& & AI & REF & AI & REF & AI & REF \\
\midrule
\endhead

\midrule
\multicolumn{8}{r@{}}{\small\itshape continued on next page} \\
\endfoot

\bottomrule
\endlastfoot

\multicolumn{8}{c}{\textbf{Bayesian nonparametrics}} \\
\midrule
\model{continuous_BART} & 1.0006 & 0.97 & 0.97 & 3562 & 4673 & 43.1 & 50.0 \\
\model{logistic_BART} & 1.0034 & 0.95 & --- & 513 & --- & 118.9 & --- \\
\model{poisson_BART} & 1.0022 & 0.97 & --- & 731 & --- & 349.5 & --- \\
\model{negative_binomial_BART} & 1.0018 & 0.97 & --- & 971 & --- & 307.2 & --- \\
\model{probit_BART} & 1.0011 & 0.97 & 0.97 & 1955 & 1950 & 7.7 & 9.0 \\
\model{DPM_Gaussian} & 1.0038 & 0.95 & 0.94 & 664 & 419 & 126.1 & 522.0 \\
\model{GP_normal} & 1.0001 & 0.96 & 0.96 & 10216 & 11988 & 4.0 & 7.6 \\
\model{gp_regr} & 1.0001 & 0.94 & 0.94 & 15004 & 17060 & 3.9 & 8.0 \\
\model{gp_pois_regr} & 1.0001 & 0.95 & 0.95 & 9699 & 8862 & 123.3 & 29.2 \\
\model{hierarchical_gp} & 1.0005 & 0.95 & 0.95 & 2552 & 8010 & 29.8 & 59.4 \\
\model{kronecker_gp} & 1.0002 & 0.93 & 0.91 & 7759 & 11384 & 2.2 & 8.9 \\
\model{accel_gp} & 1.0033 & 0.65 & 0.66 & 498 & 778 & 57.3 & 277.9 \\
\model{nn_rbm1bJ10} & 1.0004 & 0.94 & 0.93 & 5629 & 10026 & 685.5 & 76.4 \\
\model{nn_rbm1bJ100} & 1.0012 & 0.93 & 0.93 & 342 & 11864 & 685.3 & 704.4 \\
\midrule

\multicolumn{8}{c}{\textbf{Mixture / clustering}} \\
\midrule
\model{normal_mixture} & 1.0002 & 0.75 & 0.75 & 11760 & 18750 & 0.59 & 3.4 \\
\model{normal_mixture_k} & 1.0073 & 0.98 & 0.99 & 353 & 2333 & 66.8 & 102.0 \\
\model{low_dim_gauss_mix} & 1.0067 & 0.95 & 0.90 & 3858 & 6865 & 3.7 & 14.5 \\
\model{low_dim_gauss_mix_collapse} & 1.0004 & 0.93 & 0.92 & 1313 & 6554 & 2.0 & 11.9 \\
\midrule

\multicolumn{8}{c}{\textbf{Discrete latent variable models}} \\
\midrule
\model{Dirac_Spike_Laplace} & 1.0012 & 0.98 & 0.97 & 1751 & 789 & 1.1 & 80.2 \\
\model{Dirac_Spike_normal} & 1.0008 & 0.93 & 0.96 & 3105 & 1350 & 2.5 & 51.6 \\
\model{ldaK2} & 1.0004 & 0.97 & 0.97 & 2872 & 8190 & 0.37 & 37.4 \\
\model{ldaK5} & 1.0003 & 0.89 & 0.89 & 3741 & 9583 & 0.80 & 96.7 \\
\model{poisson_changepoint} & 1.0003 & 0.94 & 0.95 & 5099 & 7002 & 0.36 & 4.7 \\
\model{poisson_k_changepoint} & 1.0003 & 0.95 & 0.96 & 4618 & 6734 & 2.0 & 10.8 \\
\model{iohmm_reg} & 1.0003 & 0.94 & 0.90 & 15590 & 12837 & 7.8 & 68.9 \\
\model{M0_model} & 1.0001 & 0.96 & 0.96 & 10512 & 11628 & 0.75 & 3.2 \\
\model{Mb_model} & 1.0002 & 0.95 & 0.95 & 2984 & 10183 & 0.84 & 14.1 \\
\model{Mh_model} & 1.0043 & 0.96 & 0.97 & 3120 & 2823 & 52.5 & 13.3 \\
\model{Mt_model} & 1.0001 & 0.95 & 0.95 & 14649 & 27023 & 1.0 & 5.1 \\
\model{Mtbh_model} & 1.0012 & 0.97 & 0.97 & 4691 & 3910 & 19.0 & 19.1 \\
\model{Mth_model} & 1.0008 & 0.98 & 0.99 & 3887 & 3843 & 20.0 & 17.2 \\
\model{multi_occupancy} & 1.0078 & 0.96 & 0.92 & 633 & 233 & 77.4 & 97.3 \\
\model{Survey_model} & 1.0002 & 0.98 & 0.98 & 5380 & 3977 & 0.19 & 5.9 \\
\midrule

\multicolumn{8}{c}{\textbf{Hierarchical models}} \\
\midrule
\model{GLMM_Poisson_model} & 1.0019 & 0.98 & 0.97 & 6845 & 437 & 18.3 & 7.0 \\
\model{GLMM1_model} & 1.0055 & 0.96 & 0.96 & 2718 & 3843 & 2.4 & 2.1 \\
\model{bones_model} & 1.0005 & 0.94 & 0.95 & 4485 & 8220 & 13.7 & 142.1 \\
\model{dirichlet_meanprec_hier} & 1.0091 & 0.72 & 0.73 & 1734 & 12603 & 94.0 & 82.0 \\
\model{eight_schools_centered} & 1.0018 & 0.94 & 0.94 & 5827 & 665 & 0.99 & 6.2 \\
\model{eight_schools_noncentered} & 1.0007 & 0.96 & 0.96 & 5828 & 7593 & 0.60 & 1.4 \\
\model{election88_full} & 1.0027 & 0.97 & 0.96 & 2959 & 520 & 380.0 & 51.8 \\
\model{pilots} & 1.0004 & 0.97 & 0.96 & 5025 & 1472 & 135.7 & 85.7 \\
\model{rats_model} & 1.0053 & 0.93 & 0.95 & 312 & 11304 & 12.2 & 2.9 \\
\model{surgical_model} & 1.0042 & 0.96 & 0.96 & 5134 & 1450 & 1.4 & 1.3 \\
\model{radon_county} & 1.0070 & 0.96 & 0.96 & 3692 & 15507 & 3.6 & 2.0 \\
\model{radon_county_intercept} & 1.0004 & 0.94 & 0.94 & 14847 & 12658 & 0.63 & 3.1 \\
\model{radon_hierarchical_intercept_centered} & 1.0044 & 0.95 & 0.95 & 4272 & 14294 & 3.7 & 5.7 \\
\model{radon_hierarchical_intercept_noncentered} & 1.0041 & 0.95 & 0.95 & 4511 & 5247 & 3.9 & 6.2 \\
\model{radon_partially_pooled_centered} & 1.0003 & 0.94 & 0.95 & 13532 & 15161 & 0.90 & 3.0 \\
\model{radon_partially_pooled_noncentered} & 1.0070 & 0.95 & 0.95 & 3679 & 4591 & 1.6 & 4.5 \\
\model{radon_variable_intercept_centered} & 1.0003 & 0.94 & 0.94 & 14943 & 15424 & 1.2 & 3.9 \\
\model{radon_variable_intercept_noncentered} & 1.0042 & 0.94 & 0.94 & 4535 & 5166 & 1.6 & 5.3 \\
\model{radon_variable_intercept_slope_centered} & 1.0019 & 0.94 & 0.94 & 1201 & 1915 & 5.4 & 8.8 \\
\model{radon_variable_intercept_slope_noncentered} & 1.0015 & 0.95 & 0.94 & 6321 & 5771 & 5.8 & 9.0 \\
\model{radon_variable_slope_centered} & 1.0003 & 0.95 & 0.95 & 10505 & 12532 & 1.6 & 4.7 \\
\model{radon_variable_slope_noncentered} & 1.0024 & 0.95 & 0.95 & 4789 & 5616 & 1.9 & 5.6 \\
\model{seeds_centered_model} & 1.0014 & 0.96 & 0.96 & 6137 & 4153 & 4.1 & 5.0 \\
\model{seeds_model} & 1.0009 & 0.95 & 0.94 & 7741 & 4637 & 18.1 & 4.8 \\
\model{seeds_stanified_model} & 1.0017 & 0.97 & 0.97 & 6113 & 3755 & 4.2 & 4.2 \\
\model{pl2_latent_reg_irt} & 1.0029 & 0.94 & 0.94 & 7889 & 7090 & 107.9 & 18.4 \\
\model{gpcm_latent_reg_irt} & 1.0032 & 0.93 & 0.91 & 5024 & 3791 & 333.0 & 415.2 \\
\model{grsm_latent_reg_irt} & 1.0007 & 0.93 & 0.93 & 1823 & 3471 & 7.1 & 241.7 \\
\model{hier_2pl} & 1.0024 & 0.95 & 0.94 & 1058 & 641 & 554.9 & 79.5 \\
\model{irt_2pl} & 1.0072 & 0.97 & 0.96 & 257 & 265 & 7.7 & 56.7 \\
\model{lsat_model} & 1.0005 & 0.95 & 0.95 & 5475 & 5428 & 170.0 & 12.5 \\
\midrule

\multicolumn{8}{c}{\textbf{Time series / state-space}} \\
\midrule
\model{arK} & 1.0004 & 0.96 & 0.95 & 5333 & 8292 & 1.3 & 3.5 \\
\model{arma11} & 1.0022 & 0.98 & 0.89 & 389 & 3291 & 0.58 & 4.9 \\
\model{garch11} & 1.0092 & 0.94 & 0.93 & 5809 & 9778 & 3.6 & 18.9 \\
\model{prophet} & 1.0008 & 0.91 & 0.90 & 7679 & 7471 & 35.6 & 14.1 \\
\model{state_space_stochastic_level_stochastic_seasonal} & 1.0044 & 0.95 & 0.95 & 224 & 281 & 831.4 & 24.7 \\
\midrule

\multicolumn{8}{c}{\textbf{ODE / Mechanistic models}} \\
\midrule
\model{sir} & 1.0032 & 0.96 & 0.96 & 984 & 3478 & 187.7 & 403.8 \\
\model{covid19imperial_v2} & 1.0017 & 0.58 & 0.58 & 1040 & 1654 & 35.9 & 40.8 \\
\model{covid19imperial_v3} & 1.0099 & 0.62 & 0.59 & 357 & 374 & 137.2 & 92.6 \\
\model{lotka_volterra} & 1.0031 & 0.96 & 0.95 & 4223 & 5195 & 151.2 & 192.5 \\
\model{one_comp_mm_elim_abs} & 1.0003 & 0.91 & 0.92 & 7625 & 9606 & 5.9 & 331.1 \\
\model{soil_incubation} & 1.0040 & 0.93 & 0.92 & 4401 & 9260 & 301.3 & 1290.7 \\
\model{dugongs_model} & 1.0014 & 0.91 & 0.94 & 2566 & 5045 & 1.1 & 2.3 \\
\model{losscurve_sislob} & 1.0005 & 0.95 & 0.96 & 6808 & 7534 & 8.3 & 74.2 \\
\midrule

\multicolumn{8}{c}{\textbf{Graphical models}} \\
\midrule
\model{bym2_offset_only} & 1.0004 & 0.97 & 0.97 & 4655 & 2922 & 62.4 & 110.2 \\
\model{hmm_drive_0} & 1.0002 & 0.95 & 0.95 & 6520 & 19751 & 2.2 & 39.8 \\
\model{hmm_drive_1} & 1.0003 & 0.95 & 0.96 & 13377 & 16512 & 3.2 & 59.9 \\
\model{hmm_example} & 1.0002 & 0.93 & 0.93 & 6019 & 11133 & 0.91 & 21.1 \\
\model{hmm_gaussian} & 1.0004 & 0.96 & 0.84 & 5726 & 16889 & 2.8 & 89.0 \\
\model{hex_ising_seg_0} & 1.0018 & 0.94 & 0.93 & 3237 & 7496 & 2.3 & 4.5 \\
\model{pois_hmm_k3_0} & 1.0002 & 0.95 & 0.95 & 5499 & 16937 & 5.1 & 149.9 \\
\model{water_bif} & 1.0392 & 0.97 & 0.97 & 85 & 165 & 29.0 & 326.5 \\
\midrule

\multicolumn{8}{c}{\textbf{Regression models}} \\
\midrule
\model{blr} & 1.0001 & 0.95 & 0.95 & 12842 & 17705 & 0.32 & 0.80 \\
\model{diamonds} & 1.0002 & 0.96 & 0.96 & 17037 & 23738 & 0.32 & 0.82 \\
\model{earn_height} & 1.0008 & 0.93 & 0.93 & 6830 & 5376 & 1.7 & 4.7 \\
\model{kilpisjarvi} & 1.0017 & 0.94 & 0.94 & 13397 & 17333 & 0.30 & 0.82 \\
\model{radon_pooled} & 1.0002 & 0.97 & 0.97 & 9563 & 10809 & 0.36 & 1.9 \\
\model{kidscore_interaction} & 1.0008 & 0.96 & 0.94 & 8397 & 10681 & 1.1 & 2.0 \\
\model{kidscore_interaction_c} & 1.0019 & 0.94 & 0.93 & 14454 & 21589 & 0.55 & 1.1 \\
\model{kidscore_interaction_c2} & 1.0018 & 0.94 & 0.94 & 11856 & 13743 & 0.80 & 1.7 \\
\model{kidscore_interaction_z} & 1.0019 & 0.93 & 0.94 & 15159 & 21321 & 1.2 & 1.6 \\
\model{kidscore_mom_work} & 1.0016 & 0.95 & 0.94 & 7071 & 7417 & 1.1 & 1.7 \\
\model{kidscore_momhs} & 1.0003 & 0.94 & 0.95 & 8161 & 8401 & 0.35 & 1.3 \\
\model{kidscore_momhsiq} & 1.0002 & 0.97 & 0.97 & 14743 & 7306 & 0.38 & 2.9 \\
\model{kidscore_momiq} & 1.0049 & 0.97 & 0.96 & 2821 & 5810 & 0.98 & 2.0 \\
\model{mesquite} & 1.0011 & 0.56 & 0.56 & 5948 & 12238 & 1.1 & 7.6 \\
\model{logmesquite} & 1.0020 & 0.94 & 0.94 & 14918 & 9593 & 1.2 & 6.8 \\
\model{logmesquite_logva} & 1.0029 & 0.95 & 0.96 & 9435 & 8843 & 6.5 & 9.2 \\
\model{logmesquite_logvas} & 1.0034 & 0.94 & 0.95 & 15546 & 9644 & 13.7 & 16.9 \\
\model{logmesquite_logvash} & 1.0031 & 0.94 & 0.95 & 13751 & 9748 & 10.9 & 14.7 \\
\model{logmesquite_logvolume} & 1.0011 & 0.93 & 0.93 & 8059 & 9027 & 0.34 & 1.1 \\
\model{log10earn_height} & 1.0078 & 0.97 & 0.96 & 1727 & 5240 & 3.3 & 5.9 \\
\model{logearn_height} & 1.0015 & 0.96 & 0.96 & 7658 & 5324 & 1.5 & 3.9 \\
\model{logearn_height_male} & 1.0016 & 0.97 & 0.96 & 8487 & 7132 & 2.0 & 5.8 \\
\model{logearn_interaction} & 1.0035 & 0.97 & 0.96 & 1470 & 4575 & 18.2 & 18.2 \\
\model{logearn_interaction_z} & 1.0009 & 0.95 & 0.94 & 9784 & 11021 & 0.49 & 1.8 \\
\model{logearn_logheight_male} & 1.0041 & 0.95 & 0.95 & 583 & 6381 & 30.5 & 24.9 \\
\model{nes} & 1.0037 & 0.94 & 0.94 & 12918 & 9328 & 1.3 & 4.7 \\
\model{sesame_one_pred_a} & 1.0023 & 0.93 & 0.93 & 12787 & 16611 & 0.26 & 0.86 \\
\model{GLM_Binomial_model} & 1.0002 & 0.94 & 0.94 & 8762 & 8583 & 0.31 & 1.2 \\
\model{dogs} & 1.0002 & 0.89 & 0.93 & 11043 & 8344 & 5.8 & 9.9 \\
\model{dogs_hierarchical} & 1.0002 & 0.94 & 0.94 & 7553 & 8451 & 0.36 & 2.2 \\
\model{dogs_log} & 1.0003 & 0.94 & 0.94 & 4284 & 5262 & 0.52 & 4.3 \\
\model{dogs_nonhierarchical} & 1.0026 & 0.99 & 0.99 & 7378 & 9521 & 5.4 & 12.3 \\
\model{nes_logit_model} & 1.0004 & 0.92 & 0.95 & 3772 & 8360 & 3.1 & 0.94 \\
\model{wells_daae_c_model} & 1.0003 & 0.89 & 0.88 & 3948 & 6084 & 4.5 & 13.7 \\
\model{wells_dae_c_model} & 1.0012 & 0.79 & 0.78 & 5558 & 6157 & 1.2 & 1.7 \\
\model{wells_dae_inter_model} & 1.0016 & 0.73 & 0.71 & 5127 & 4944 & 1.7 & 2.0 \\
\model{wells_dae_model} & 1.0002 & 0.96 & 0.97 & 11510 & 9660 & 71.1 & 9.3 \\
\model{wells_dist} & 1.0008 & 0.91 & 0.91 & 3605 & 5142 & 0.69 & 1.4 \\
\model{wells_dist100_model} & 1.0003 & 0.95 & 0.95 & 4091 & 5369 & 0.60 & 1.2 \\
\model{wells_dist100ars_model} & 1.0006 & 0.85 & 0.85 & 4782 & 6085 & 0.85 & 1.8 \\
\model{wells_interaction_c_model} & 1.0003 & 0.88 & 0.87 & 6331 & 7472 & 0.78 & 1.4 \\
\model{wells_interaction_model} & 1.0090 & 0.78 & 0.76 & 1885 & 3431 & 3.6 & 2.8 \\
\model{Rate_1_model} & 1.0001 & 0.97 & 0.97 & 19867 & 6181 & 0.01 & 0.37 \\
\model{Rate_2_model} & 1.0001 & 0.95 & 0.95 & 19568 & 12882 & 0.03 & 0.49 \\
\model{Rate_3_model} & 1.0001 & 0.95 & 0.95 & 19898 & 6399 & 0.02 & 0.38 \\
\model{Rate_4_model} & 1.0001 & 0.97 & 0.97 & 19578 & 12314 & 0.02 & 0.50 \\
\model{Rate_5_model} & 1.0001 & 0.95 & 0.96 & 19898 & 6377 & 0.02 & 0.48 \\
\model{GLM_Poisson_model} & 1.0013 & 0.94 & 0.94 & 4741 & 6471 & 0.81 & 2.0 \\
\model{two_normal_spike_slab_model} & 1.0008 & 0.94 & 0.95 & 2065 & 12172 & 0.66 & 2.8 \\
\model{logistic_regression_rhs} & 1.0003 & 0.94 & 0.94 & 4233 & 9989 & 0.98 & 5.4 \\
\model{accel_splines} & 1.0038 & 0.91 & 0.93 & 281 & 3154 & 5.6 & 42.8 \\
\end{longtable}
\endgroup

\newpage
\section{Results of Experiment 1 generated by Codex}
To illustrate the usage of our system on a different platform, we conducted Experiment 1 using Codex GPT-5.5 with Extra High reasoning, in place of Claude Code. For this demonstration, we restricted attention to the 16 models used in Experiment 2. The detailed results are summarized in Web Table~\ref{tab:sim1_codex_results}.

\begingroup
\scriptsize
\setlength{\tabcolsep}{3pt}
\renewcommand{\arraystretch}{1.03}
\setlength{\LTcapwidth}{\linewidth}
\begin{longtable}{@{}L{4.0cm} C{1.15cm} C{1.3cm} C{0.8cm} C{0.8cm} C{0.95cm} C{0.95cm} C{0.95cm} C{0.95cm}@{}}
\caption{Results for the OpenAI Codex GPT-5.5 with Extra High reasoning, grouped by category. This table reports the 16 models for which we obtained Codex implementations. For each benchmark model we summarize 100 replicate datasets by reporting the median of the maximum rank-normalized $\hat{R}$ over all parameters, the mean coverage of 95\% credible intervals across parameters, the median of the minimum bulk effective sample size, and the median running time. The generalized BART model (\texttt{negative\_binomial\_BART}) has no reference implementation and reports ``---'' in the reference columns.}
\label{tab:sim1_codex_results}\\
\multirow{2}{*}{\textbf{Model}} &
\multirow{2}{*}{\textbf{Status}} &
\multirow{2}{*}{\shortstack[c]{\textbf{Median}\\\textbf{max} $\boldsymbol{\hat{R}}$}} &
\multicolumn{2}{c}{\textbf{Mean coverage}} &
\multicolumn{2}{c}{\textbf{Min ESS}} &
\multicolumn{2}{c}{\textbf{Runtime (s)}} \\
\cmidrule(lr){4-5} \cmidrule(lr){6-7} \cmidrule(lr){8-9}
& & & AI & REF & AI & REF & AI & REF \\
\midrule
\endfirsthead
\multicolumn{9}{l}{\emph{Continued from previous page}} \\
\multirow{2}{*}{\textbf{Model}} &
\multirow{2}{*}{\textbf{Status}} &
\multirow{2}{*}{\shortstack[c]{\textbf{Median}\\\textbf{max} $\boldsymbol{\hat{R}}$}} &
\multicolumn{2}{c}{\textbf{Mean coverage}} &
\multicolumn{2}{c}{\textbf{Min ESS}} &
\multicolumn{2}{c}{\textbf{Runtime (s)}} \\
\cmidrule(lr){4-5} \cmidrule(lr){6-7} \cmidrule(lr){8-9}
& & & AI & REF & AI & REF & AI & REF \\
\midrule
\endhead
\multicolumn{9}{r}{\emph{Continued on next page}} \\
\endfoot
\vspace{\baselineskip}
\endlastfoot

\multicolumn{9}{c}{\textbf{Bayesian nonparametrics}} \\
\midrule
\model{DPM_Gaussian} & Success & 1.005 & 0.95 & 0.94 & 434 & 419 & 305.5 & 593.4 \\
\model{gp_pois_regr} & Success & 1.001 & 0.80 & 0.95 & 917 & 7519 & 9.25 & 30.2 \\
\model{negative_binomial_BART} & Success & 1.002 & 0.97 & --- & 466 & --- & 185.6 & --- \\
\model{probit_BART} & Success & 1.001 & 0.97 & 0.97 & 1955 & 1950 & 9.02 & 10.4 \\
\midrule

\multicolumn{9}{c}{\textbf{Mixture / clustering}} \\
\midrule
\model{normal_mixture} & Success & 1.002 & 0.72 & 0.75 & 3258 & 18959 & 1.35 & 2.51 \\
\midrule

\multicolumn{9}{c}{\textbf{Discrete latent variable models}} \\
\midrule
\model{Dirac_Spike_Laplace} & Success & 1.008 & 0.97 & 0.98 & 185 & 728 & 3.60 & 71.6 \\
\model{poisson_k_changepoint} & Success & 1.000 & 0.96 & 0.96 & 3873 & 7444 & 14.7 & 7.57 \\
\midrule

\multicolumn{9}{c}{\textbf{Hierarchical models}} \\
\midrule
\model{GLMM1_model} & Success & 1.006 & 0.96 & 0.96 & 1862 & 4044 & 3.80 & 1.58 \\
\model{irt_2pl} & Success & 1.022 & 0.95 & 0.97 & 131 & 200 & 661.1 & 21.8 \\
\midrule

\multicolumn{9}{c}{\textbf{Regression models}} \\
\midrule
\model{dogs_log} & Success & 1.000 & 0.94 & 0.94 & 3017 & 5606 & 1.30 & 3.34 \\
\model{wells_daae_c_model} & Success & 1.004 & 0.79 & 0.76 & 5186 & 6815 & 8.47 & 2.46 \\
\model{logistic_regression_rhs} & Success & 1.000 & 0.94 & 0.94 & 13753 & 13457 & 5.35 & 3.34 \\
\midrule

\multicolumn{9}{c}{\textbf{Time series / state-space}} \\
\midrule
\model{arma11} & Success & 1.010 & 0.98 & 0.93 & 255 & 2292 & 1.78 & 2.76 \\
\model{garch11} & Success & 1.003 & 0.90 & 0.90 & 943 & 8972 & 3.46 & 2.05 \\
\midrule

\multicolumn{9}{c}{\textbf{ODE / Mechanistic models}} \\
\midrule
\model{one_comp_mm_elim_abs} & Success & 1.002 & 0.86 & 0.90 & 3794 & 10666 & 134.8 & 297.5 \\
\midrule

\multicolumn{9}{c}{\textbf{Graphical models}} \\
\midrule
\model{hmm_drive_0} & Success & 1.000 & 0.95 & 0.95 & 4692 & 22665 & 2.54 & 30.5 \\
\midrule

\end{longtable}
\endgroup

\newpage
\raggedbottom
\section{Results of Experiment 1 generated by direct LLM prompts}

To evaluate sampler generation without AI4BayesCode, we conducted a direct prompting experiment using the LLM alone. The prompts for the \texttt{R} and \texttt{Python} interfaces are specified in Prompt~\ref{prompt:plain-r} and Prompt~\ref{prompt:plain-python}, respectively. We generated samplers using Claude Code Opus 4.7 (max effort) for a fair comparison with Experiment~1, and Claude Code Opus 5 (max effort), the newer Opus model, to evaluate direct prompting with the most recent model. For each benchmark model, we replaced the placeholder \texttt{\{Natural-language model description\}} with the full natural-language description of that model and asked the LLM to generate a self-contained C++ MCMC sampler, exposed either to \texttt{R} or to \texttt{Python}. We used the same evaluation settings and criteria as in Experiment~1 of the main paper. However, direct AI-generated samplers did not always return the same set of parameters as AI4BayesCode samplers. Therefore, comparisons with the reference implementation were restricted to the parameters returned by each generated sampler, and the corresponding reference results may differ slightly from those reported in Experiment~1.

The results for Opus 4.7 are summarized in Web Table~\ref{tab:sim1_results_all_opus47}. Direct prompting failed to generate runnable samplers for two models, \texttt{normal\_mixture\_k} and \texttt{wells\_dist100ars\_model}. Among the runnable samplers, Opus 4.7 performed well for many simple and classical models, but performance became substantially less reliable for more challenging models. In total, 23 models had $\hat R \geq 1.01$, including 11 with $\hat R \geq 1.05$, with particularly poor performance for complex models such as \texttt{nn\_rbm1bJ10}, \texttt{nn\_rbm1bJ100}, \texttt{bym2\_offset\_only}, \texttt{pois\_hmm\_k3\_0}, and \texttt{water\_bif}. For nonconjugate models, Opus 4.7 often relied on simple Metropolis--Hastings updates and therefore produced low ESS, in some cases below 100.

Direct prompting improved substantially with Opus 5 (see Web Table~\ref{tab:sim1_results_all_opus5}), reducing the number of models with $\hat R \geq 1.01$ from 23 to 7. However, it continued to struggle with several challenging models, including \texttt{continuous\_BART}, \texttt{nn\_rbm1bJ100}, and \texttt{water\_bif}. In contrast, the modular design of AI4BayesCode provides validated specialized sampling blocks for such complex models, yielding substantially more stable sampler generation. Opus 5 also more frequently implemented NUTS for nonconjugate models, which substantially improved mixing and ESS relative to Opus 4.7. Notably, there is still a clear gap relative to AI4BayesCode, despite AI4BayesCode itself using only Opus 4.7 in Supplementary Section~\ref{supp:exp1}.

In addition, direct prompting samplers were usually self-contained, with fewer function calls, variable initializations, and cached quantities, which explains their faster running time. However, they lacked a stateful sampler design. Consequently, embedding such code into a larger MCMC algorithm would generally require rewriting substantial portions of the code, increasing the risk of implementation errors. 

\begin{promptbox}[label={prompt:plain-r}]{Direct prompting baseline prompt --- R}
\begin{Verbatim}[
  breaklines=true,
  breakanywhere=false,
  breaksymbolleft={},
  breaksymbolright={},
  breakindent=0pt
]
You will implement a Bayesian model in C++ from scratch, exposed to R via Rcpp. Use Jeffreys prior if a prior is not specified.

TASK
----
Produce a single .cpp file that compiles via Rcpp::sourceCpp and exposes:

  Rcpp::List run_mcmc(Rcpp::List data, int n_burnin, int n_keep, int seed)

returning a named Rcpp::List of posterior draws (scalar parameter -> NumericVector of length n_keep; vector parameter -> NumericMatrix [n_keep, dim]; element names = parameter names from the spec). End the generation immediately once it compiles successfully. If compilation fails, revise the code until it can compile successfully. Allow up to 5 retries.

CONSTRAINTS
-----------
C++ only (Rcpp + RcppArmadillo + STL). Do NOT link Stan / JAGS / NIMBLE / PyMC headers. You must implement your own MCMC kernel.

MODEL SPECIFICATION
-------------------
{Natural-language model description}
\end{Verbatim}
\end{promptbox}

\begin{promptbox}[label={prompt:plain-python}]{Direct prompting baseline prompt --- Python}
\begin{Verbatim}[
  breaklines=true,
  breakanywhere=false,
  breaksymbolleft={},
  breaksymbolright={},
  breakindent=0pt
]
You will implement a Bayesian model in C++ from scratch, callable from Python via pybind11. Use Jeffreys prior if a prior is not specified.

TASK
----
Produce a single .cpp file that compiles via cppimport.imp_from_filepath(...) and exposes:

  py::dict run_mcmc(py::dict data, int n_burnin, int n_keep, int seed)

returning a Python dict of posterior draws (scalar parameter -> np.ndarray shape (n_keep,); vector parameter -> np.ndarray shape (n_keep, dim); keys = parameter names from the spec). End the generation immediately once it compiles successfully. If compilation fails, revise the code until it can compile successfully. Allow up to 5 retries.

CONSTRAINTS
-----------
C++ only (pybind11 + Eigen + STL). Do NOT link Stan / JAGS / NIMBLE / PyMC headers. You must implement your own MCMC kernel.

MODEL SPECIFICATION
-------------------
{Natural-language model description}
\end{Verbatim}
\end{promptbox}

\begingroup
\fontsize{9.0}{9.6}\selectfont
\renewcommand{\arraystretch}{0.86}
\setlength{\tabcolsep}{3pt}

\setlength{\LTcapwidth}{\textwidth}
\captionsetup[longtable]{
    justification=raggedright,
    singlelinecheck=false
}

\begin{longtable}{@{}
    >{\raggedright\arraybackslash}p{6.7cm}
    >{\centering\arraybackslash}p{1.35cm}
    >{\centering\arraybackslash}p{1.05cm}
    >{\centering\arraybackslash}p{1.05cm}
    >{\centering\arraybackslash}p{1.15cm}
    >{\centering\arraybackslash}p{1.15cm}
    >{\centering\arraybackslash}p{1.15cm}
    >{\centering\arraybackslash}p{1.15cm}
    @{}}

\caption{Per-model results for direct AI-generated samplers in Experiment 1 using Claude Code \textbf{Opus 4.7} (max effort). For each benchmark model we summarize 100 replicate datasets by reporting the median of the maximum rank-normalized $\hat{R}$ over all parameters, the mean coverage of 95\% credible intervals across parameters, the median of the minimum bulk effective sample size, and the median running time.}
\label{tab:sim1_results_all_opus47} \\

\toprule
\multirow{2}{*}{\textbf{Model}} &
\multirow{2}{*}{\shortstack[c]{\textbf{Median}\\\textbf{max} $\boldsymbol{\hat{R}}$}} &
\multicolumn{2}{c}{\textbf{Mean coverage}} &
\multicolumn{2}{c}{\textbf{Min ESS}} &
\multicolumn{2}{c}{\textbf{Runtime (s)}} \\
\cmidrule(lr){3-4}
\cmidrule(lr){5-6}
\cmidrule(lr){7-8}
& & AI & REF & AI & REF & AI & REF \\
\midrule
\endfirsthead

\multicolumn{8}{@{}l@{}}{\small\itshape
(Table~\thetable\ continued from previous page)} \\
\toprule
\multirow{2}{*}{\textbf{Model}} &
\multirow{2}{*}{\shortstack[c]{\textbf{Median}\\\textbf{max} $\boldsymbol{\hat{R}}$}} &
\multicolumn{2}{c}{\textbf{Mean coverage}} &
\multicolumn{2}{c}{\textbf{Min ESS}} &
\multicolumn{2}{c}{\textbf{Runtime (s)}} \\
\cmidrule(lr){3-4}
\cmidrule(lr){5-6}
\cmidrule(lr){7-8}
& & AI & REF & AI & REF & AI & REF \\
\midrule
\endhead

\midrule
\multicolumn{8}{r@{}}{\small\itshape continued on next page} \\
\endfoot

\bottomrule
\endlastfoot

\multicolumn{8}{c}{\textbf{Bayesian nonparametrics}} \\
\midrule
\model{continuous_BART} & 1.0688 & 0.96 & 0.97 & 2730 & 4673 & 11.3 & 58.0 \\
\model{logistic_BART} & 1.0011 & 0.98 & --- & 4369 & --- & 27.2 & --- \\
\model{poisson_BART} & 1.0002 & 0.98 & --- & 23682 & --- & 26.8 & --- \\
\model{negative_binomial_BART} & 1.0023 & 0.97 & --- & 743 & --- & 14.3 & --- \\
\model{probit_BART} & 1.0429 & 0.98 & 0.97 & 1681 & 1950 & 3.4 & 9.3 \\
\model{DPM_Gaussian} & 1.0060 & 0.95 & 0.94 & 543 & 419 & 0.90 & 533.2 \\
\model{GP_normal} & 1.0003 & 0.96 & 0.96 & 3253 & 11988 & 0.83 & 6.6 \\
\model{gp_regr} & 1.0006 & 0.94 & 0.94 & 1637 & 17060 & 0.25 & 7.8 \\
\model{gp_pois_regr} & 1.0027 & 0.86 & 0.95 & 182 & 8862 & 0.95 & 34.0 \\
\model{hierarchical_gp} & 1.0454 & 0.95 & 0.96 & 12 & 5616 & 15.2 & 54.0 \\
\model{kronecker_gp} & 1.0005 & 0.92 & 0.91 & 1945 & 11384 & 0.64 & 9.8 \\
\model{accel_gp} & 1.0296 & 0.67 & 0.66 & 19 & 778 & 1.4 & 271.5 \\
\model{nn_rbm1bJ10} & 1.6567 & 0.11 & 0.93 & 4 & 10026 & 8.5 & 62.5 \\
\model{nn_rbm1bJ100} & 1.3389 & 0.53 & 0.94 & 2 & 11982 & 43.0 & 546.1 \\
\midrule

\multicolumn{8}{c}{\textbf{Mixture / clustering}} \\
\midrule
\model{normal_mixture} & 1.0001 & 0.74 & 0.75 & 19184 & 18750 & 0.07 & 3.2 \\
\model{normal_mixture_k} & \multicolumn{7}{c}{\textbf{failed to generate a runnable sampler}} \\
\model{low_dim_gauss_mix} & 1.0009 & 0.93 & 0.90 & 810 & 6865 & 0.10 & 16.8 \\
\model{low_dim_gauss_mix_collapse} & 1.0007 & 0.92 & 0.92 & 835 & 6554 & 0.11 & 10.1 \\
\midrule

\multicolumn{8}{c}{\textbf{Discrete latent variable models}} \\
\midrule
\model{Dirac_Spike_Laplace} & 1.0009 & 0.97 & 0.97 & 2273 & 698 & 0.11 & 81.7 \\
\model{Dirac_Spike_normal} & 1.0006 & 0.95 & 0.96 & 10192 & 1405 & 5.6 & 49.9 \\
\model{ldaK2} & 1.0007 & 0.97 & 0.97 & 1343 & 8198 & 0.22 & 38.5 \\
\model{ldaK5} & 1.0003 & 0.73 & 0.73 & 4464 & 9590 & 0.28 & 101.7 \\
\model{poisson_changepoint} & 1.0002 & 0.96 & 0.95 & 7844 & 7002 & 0.02 & 8.2 \\
\model{poisson_k_changepoint} & 1.0004 & 0.95 & 0.96 & 2752 & 6734 & 0.35 & 9.8 \\
\model{iohmm_reg} & 1.0057 & 0.94 & 0.90 & 164 & 12837 & 1.2 & 73.8 \\
\model{M0_model} & 1.0001 & 0.96 & 0.96 & 10400 & 11628 & 0.06 & 2.5 \\
\model{Mb_model} & 1.0011 & 0.95 & 0.95 & 2622 & 10183 & 0.04 & 11.8 \\
\model{Mh_model} & 1.0083 & 0.96 & 0.97 & 59 & 2823 & 0.24 & 8.4 \\
\model{Mt_model} & 1.0013 & 0.95 & 0.95 & 14615 & 27023 & 0.06 & 5.7 \\
\model{Mtbh_model} & 1.0063 & 0.96 & 0.97 & 63 & 3910 & 0.52 & 21.1 \\
\model{Mth_model} & 1.2446 & 0.95 & 0.99 & 11 & 3843 & 0.50 & 18.8 \\
\model{multi_occupancy} & 1.1221 & 0.94 & 0.95 & 10 & 229 & 0.26 & 126.3 \\
\model{Survey_model} & 1.0006 & 0.98 & 0.98 & 830 & 3977 & 0.20 & 6.7 \\
\midrule

\multicolumn{8}{c}{\textbf{Hierarchical models}} \\
\midrule
\model{GLMM_Poisson_model} & 1.0142 & 0.97 & 0.97 & 46 & 437 & 0.07 & 6.0 \\
\model{GLMM1_model} & 1.0011 & 0.96 & 0.96 & 651 & 3843 & 0.04 & 2.9 \\
\model{bones_model} & 1.0004 & 0.94 & 0.95 & 5443 & 8220 & 3.9 & 142.1 \\
\model{dirichlet_meanprec_hier} & 1.8266 & 0.65 & 0.73 & 2 & 12603 & 0.16 & 90.7 \\
\model{eight_schools_centered} & 1.0030 & 0.94 & 0.94 & 237 & 665 & 0.01 & 6.1 \\
\model{eight_schools_noncentered} & 1.0003 & 0.94 & 0.94 & 1725 & 7593 & 0.02 & 1.7 \\
\model{election88_full} & 1.0093 & 0.97 & 0.96 & 76 & 520 & 10.1 & 45.4 \\
\model{pilots} & 1.2566 & 0.44 & 0.96 & 2 & 1492 & 0.11 & 74.1 \\
\model{rats_model} & 1.0003 & 0.95 & 0.95 & 6419 & 11304 & 0.04 & 4.8 \\
\model{surgical_model} & 1.0017 & 0.95 & 0.96 & 378 & 1450 & 0.05 & 1.6 \\
\model{radon_county} & 1.0002 & 0.96 & 0.96 & 9808 & 15507 & 0.08 & 4.0 \\
\model{radon_county_intercept} & 1.0002 & 0.94 & 0.94 & 3856 & 12658 & 0.03 & 4.9 \\
\model{radon_hierarchical_intercept_centered} & 1.0005 & 0.95 & 0.95 & 2344 & 14294 & 0.33 & 9.2 \\
\model{radon_hierarchical_intercept_noncentered} & 1.0005 & 0.95 & 0.95 & 1307 & 5247 & 0.07 & 5.1 \\
\model{radon_partially_pooled_centered} & 1.0002 & 0.96 & 0.95 & 8736 & 15161 & 0.21 & 7.1 \\
\model{radon_partially_pooled_noncentered} & 1.0008 & 0.95 & 0.95 & 789 & 4591 & 0.03 & 7.2 \\
\model{radon_variable_intercept_centered} & 1.0003 & 0.94 & 0.94 & 3033 & 15424 & 0.04 & 4.5 \\
\model{radon_variable_intercept_noncentered} & 1.0005 & 0.94 & 0.94 & 1496 & 5166 & 0.03 & 4.6 \\
\model{radon_variable_intercept_slope_centered} & 1.0017 & 0.94 & 0.94 & 295 & 1915 & 0.16 & 12.8 \\
\model{radon_variable_intercept_slope_noncentered} & 1.0010 & 0.94 & 0.94 & 285 & 5771 & 0.36 & 12.5 \\
\model{radon_variable_slope_centered} & 1.0005 & 0.95 & 0.95 & 2044 & 12532 & 0.13 & 6.0 \\
\model{radon_variable_slope_noncentered} & 1.0006 & 0.95 & 0.95 & 1104 & 5616 & 0.03 & 4.9 \\
\model{seeds_centered_model} & 1.0074 & 0.96 & 0.96 & 92 & 4085 & 2.6 & 5.9 \\
\model{seeds_model} & 1.0018 & 0.95 & 0.94 & 476 & 4637 & 1.0 & 6.3 \\
\model{seeds_stanified_model} & 1.0036 & 0.97 & 0.97 & 182 & 3755 & 0.15 & 4.0 \\
\model{pl2_latent_reg_irt} & 1.0042 & 0.94 & 0.94 & 479 & 7090 & 2.7 & 27.4 \\
\model{gpcm_latent_reg_irt} & 1.0239 & 0.94 & 0.94 & 148 & 4047 & 2.6 & 248.5 \\
\model{grsm_latent_reg_irt} & 1.0146 & 0.92 & 0.94 & 150 & 4001 & 3.7 & 264.3 \\
\model{hier_2pl} & 1.0150 & 0.96 & 0.94 & 52 & 641 & 0.78 & 67.5 \\
\model{irt_2pl} & 1.0356 & 0.96 & 0.96 & 23 & 265 & 1.1 & 57.9 \\
\model{lsat_model} & 1.0010 & 0.95 & 0.95 & 726 & 5428 & 1.3 & 10.5 \\
\midrule

\multicolumn{8}{c}{\textbf{Time series / state-space}} \\
\midrule
\model{arK} & 1.0001 & 0.96 & 0.95 & 16091 & 8292 & 0.03 & 3.6 \\
\model{arma11} & 1.0113 & 0.86 & 0.89 & 58 & 3291 & 0.07 & 4.4 \\
\model{garch11} & 1.0099 & 0.94 & 0.93 & 513 & 9778 & 0.53 & 18.8 \\
\model{prophet} & 1.0094 & 0.92 & 0.90 & 132 & 7471 & 0.45 & 13.5 \\
\model{state_space_stochastic_level_stochastic_seasonal} & 1.0469 & 0.94 & 0.95 & 12 & 281 & 0.19 & 23.3  \\
\midrule

\multicolumn{8}{c}{\textbf{ODE / Mechanistic models}} \\
\midrule
\model{sir} & 1.0033 & 0.96 & 0.96 & 259 & 3478 & 0.80 & 420.9 \\
\model{covid19imperial_v2} & 1.0033 & 0.66 & 0.66 & 272 & 1654 & 1.2 & 42.4 \\
\model{covid19imperial_v3} & 1.0184 & 0.62 & 0.59 & 32 & 374 & 3.2 & 98.7 \\
\model{lotka_volterra} & 1.0024 & 0.95 & 0.94 & 231 & 7871 & 0.29 & 197.5 \\
\model{one_comp_mm_elim_abs} & 1.0006 & 0.92 & 0.91 & 3080 & 9580 & 62.1 & 357.2 \\
\model{soil_incubation} & 1.0011 & 0.94 & 0.92 & 2977 & 9260 & 0.51 & 1339.5 \\
\model{dugongs_model} & 1.0019 & 0.93 & 0.94 & 616 & 5045 & 0.03 & 2.6 \\
\model{losscurve_sislob} & 1.0010 & 0.96 & 0.96 & 898 & 7534 & 0.17 & 70.8 \\
\midrule

\multicolumn{8}{c}{\textbf{Graphical models}} \\
\midrule
\model{bym2_offset_only} & 1.4008 & 0.82 & 0.97 & 1 & 2839 & 0.08 & 88.3 \\
\model{hmm_drive_0} & 1.0001 & 0.95 & 0.95 & 6042 & 19751 & 0.67 & 34.2 \\
\model{hmm_drive_1} & 1.0001 & 0.97 & 0.97 & 18912 & 16512 & 0.41 & 58.3 \\
\model{hmm_example} & 1.0001 & 0.94 & 0.95 & 19120 & 11133 & 1.3 & 22.2 \\
\model{hmm_gaussian} & 1.0001 & 0.97 & 0.85 & 15042 & 16889 & 1.6 & 75.2 \\
\model{hex_ising_seg_0} & 1.0018 & 0.94 & 0.93 & 3157 & 7496 & 0.82 & 4.3 \\
\model{pois_hmm_k3_0} & 1.5339 & 0.94 & 0.92 & 12011 & 16937 & 5.7 & 148.2 \\
\model{water_bif} & 1.3106 & 0.96 & 0.97 & 40 & 165 & 0.20 & 256.2 \\
\midrule

\multicolumn{8}{c}{\textbf{Regression models}} \\
\midrule
\model{blr} & 1.0002 & 0.96 & 0.95 & 3915 & 17705 & 0.03 & 0.78 \\
\model{diamonds} & 1.0001 & 0.96 & 0.96 & 15501 & 23738 & 0.18 & 1.3 \\
\model{earn_height} & 1.0011 & 0.93 & 0.93 & 18928 & 5376 & 0.01 & 5.2 \\
\model{kilpisjarvi} & 1.0018 & 0.94 & 0.94 & 18265 & 17333 & 0.02 & 0.78 \\
\model{radon_pooled} & 1.0001 & 0.97 & 0.97 & 4356 & 10809 & 0.20 & 2.9 \\
\model{kidscore_interaction} & 1.0002 & 0.95 & 0.94 & 3963 & 10681 & 0.01 & 1.6 \\
\model{kidscore_interaction_c} & 1.0019 & 0.93 & 0.93 & 16943 & 21589 & 0.01 & 1.1 \\
\model{kidscore_interaction_c2} & 1.0020 & 0.94 & 0.94 & 19404 & 13743 & 0.00 & 1.2 \\
\model{kidscore_interaction_z} & 1.0018 & 0.93 & 0.94 & 19243 & 21321 & 0.01 & 1.1 \\
\model{kidscore_mom_work} & 1.0019 & 0.93 & 0.94 & 16895 & 7417 & 0.01 & 1.4 \\
\model{kidscore_momhs} & 1.0002 & 0.95 & 0.95 & 4214 & 8401 & 0.02 & 1.1 \\
\model{kidscore_momhsiq} & 1.0001 & 0.97 & 0.97 & 17641 & 7306 & 0.01 & 3.3 \\
\model{kidscore_momiq} & 1.0002 & 0.96 & 0.96 & 4238 & 5810 & 0.01 & 2.3 \\
\model{mesquite} & 1.0011 & 0.56 & 0.56 & 19155 & 12238 & 0.03 & 6.2 \\
\model{logmesquite} & 1.0011 & 0.94 & 0.94 & 19152 & 9593 & 0.01 & 6.6 \\
\model{logmesquite_logva} & 1.0012 & 0.96 & 0.96 & 17959 & 8843 & 0.02 & 11.3 \\
\model{logmesquite_logvas} & 1.0012 & 0.95 & 0.95 & 16666 & 9644 & 0.03 & 21.9 \\
\model{logmesquite_logvash} & 1.0012 & 0.95 & 0.95 & 19208 & 9748 & 0.02 & 21.8 \\
\model{logmesquite_logvolume} & 1.0018 & 0.93 & 0.93 & 18312 & 9027 & 0.04 & 2.2 \\
\model{log10earn_height} & 1.0018 & 0.96 & 0.96 & 18330 & 5240 & 0.02 & 8.0 \\
\model{logearn_height} & 1.0017 & 0.96 & 0.96 & 19571 & 5324 & 0.00 & 4.5 \\
\model{logearn_height_male} & 1.0019 & 0.96 & 0.96 & 17524 & 7132 & 0.01 & 6.2 \\
\model{logearn_interaction} & 1.0020 & 0.96 & 0.96 & 16943 & 4575 & 0.01 & 14.9 \\
\model{logearn_interaction_z} & 1.0019 & 0.94 & 0.94 & 19366 & 11021 & 0.01 & 1.5 \\
\model{logearn_logheight_male} & 1.0018 & 0.95 & 0.95 & 19479 & 6381 & 0.01 & 22.6 \\
\model{nes} & 1.0017 & 0.94 & 0.94 & 19054 & 9328 & 0.05 & 6.6 \\
\model{sesame_one_pred_a} & 1.0023 & 0.93 & 0.93 & 17974 & 16611 & 0.02 & 1.0 \\
\model{GLM_Binomial_model} & 1.0005 & 0.94 & 0.94 & 1779 & 8583 & 0.01 & 1.2 \\
\model{dogs} & 1.0008 & 0.93 & 0.93 & 794 & 8344 & 0.13 & 9.2 \\
\model{dogs_hierarchical} & 1.6614 & 0.36 & 0.94 & 1522 & 8451 & 0.01 & 2.3 \\
\model{dogs_log} & 1.0004 & 0.94 & 0.94 & 2062 & 5262 & 0.10 & 6.6 \\
\model{dogs_nonhierarchical} & 1.0248 & 0.98 & 0.99 & 32 & 9521 & 0.25 & 16.7 \\
\model{nes_logit_model} & 1.0005 & 0.95 & 0.95 & 2137 & 8360 & 0.02 & 0.83 \\
\model{wells_daae_c_model} & 1.0014 & 0.89 & 0.88 & 753 & 6084 & 0.13 & 11.4 \\
\model{wells_dae_c_model} & 1.0011 & 0.77 & 0.78 & 893 & 6157 & 0.05 & 1.7 \\
\model{wells_dae_inter_model} & 1.0018 & 0.72 & 0.71 & 562 & 4944 & 0.04 & 2.0 \\
\model{wells_dae_model} & 1.0006 & 0.96 & 0.97 & 1370 & 9660 & 0.56 & 7.0 \\
\model{wells_dist} & 1.0005 & 0.91 & 0.91 & 2297 & 5142 & 0.04 & 1.0  \\
\model{wells_dist100_model} & 1.0006 & 0.96 & 0.95 & 947 & 5369 & 0.11 & 1.1 \\
\model{wells_dist100ars_model} & \multicolumn{7}{c}{\textbf{failed to generate a runnable sampler}} \\
\model{wells_interaction_c_model} & 1.0008 & 0.87 & 0.87 & 1185 & 7472 & 0.04 & 1.2 \\
\model{wells_interaction_model} & 1.0011 & 0.75 & 0.76 & 1051 & 3431 & 0.06 & 2.4 \\
\model{Rate_1_model} & 1.0001 & 0.97 & 0.97 & 19882 & 6181 & 0.00 & 0.47 \\
\model{Rate_2_model} & 1.0001 & 0.95 & 0.95 & 19577 & 12852 & 0.01 & 0.57 \\
\model{Rate_3_model} & 1.0002 & 0.95 & 0.95 & 4466 & 6399 & 0.00 & 0.86 \\
\model{Rate_4_model} & 1.0001 & 0.97 & 0.97 & 19517 & 12314 & 0.01 & 0.47 \\
\model{Rate_5_model} & 1.0001 & 0.95 & 0.96 & 19898 & 6377 & 0.00 & 0.46 \\
\model{GLM_Poisson_model} & 1.0009 & 0.93 & 0.94 & 1181 & 6471 & 0.01 & 2.1 \\
\model{two_normal_spike_slab_model} & 1.0041 & 0.95 & 0.95 & 4613 & 12172 & 0.18 & 2.6 \\
\model{logistic_regression_rhs} & 1.0010 & 0.95 & 0.94 & 895 & 9989 & 10.1 & 5.5 \\
\model{accel_splines} & 1.0068 & 0.93 & 0.93 & 136 & 3154 & 0.33 & 41.1 \\
\midrule
\end{longtable}
\endgroup

\begingroup
\fontsize{9.0}{9.6}\selectfont
\renewcommand{\arraystretch}{0.86}
\setlength{\tabcolsep}{3pt}

\setlength{\LTcapwidth}{\textwidth}
\captionsetup[longtable]{
    justification=raggedright,
    singlelinecheck=false
}

\begin{longtable}{@{}
    >{\raggedright\arraybackslash}p{6.7cm}
    >{\centering\arraybackslash}p{1.35cm}
    >{\centering\arraybackslash}p{1.05cm}
    >{\centering\arraybackslash}p{1.05cm}
    >{\centering\arraybackslash}p{1.15cm}
    >{\centering\arraybackslash}p{1.15cm}
    >{\centering\arraybackslash}p{1.15cm}
    >{\centering\arraybackslash}p{1.15cm}
    @{}}

\caption{Per-model results for direct AI-generated samplers in Experiment 1 using Claude Code \textbf{Opus 5} (max effort). For each benchmark model we summarize 100 replicate datasets by reporting the median of the maximum rank-normalized $\hat{R}$ over all parameters, the mean coverage of 95\% credible intervals across parameters, the median of the minimum bulk effective sample size, and the median running time.}
\label{tab:sim1_results_all_opus5} \\

\toprule
\multirow{2}{*}{\textbf{Model}} &
\multirow{2}{*}{\shortstack[c]{\textbf{Median}\\\textbf{max} $\boldsymbol{\hat{R}}$}} &
\multicolumn{2}{c}{\textbf{Mean coverage}} &
\multicolumn{2}{c}{\textbf{Min ESS}} &
\multicolumn{2}{c}{\textbf{Runtime (s)}} \\
\cmidrule(lr){3-4}
\cmidrule(lr){5-6}
\cmidrule(lr){7-8}
& & AI & REF & AI & REF & AI & REF \\
\midrule
\endfirsthead

\multicolumn{8}{@{}l@{}}{\small\itshape
(Table~\thetable\ continued from previous page)} \\
\toprule
\multirow{2}{*}{\textbf{Model}} &
\multirow{2}{*}{\shortstack[c]{\textbf{Median}\\\textbf{max} $\boldsymbol{\hat{R}}$}} &
\multicolumn{2}{c}{\textbf{Mean coverage}} &
\multicolumn{2}{c}{\textbf{Min ESS}} &
\multicolumn{2}{c}{\textbf{Runtime (s)}} \\
\cmidrule(lr){3-4}
\cmidrule(lr){5-6}
\cmidrule(lr){7-8}
& & AI & REF & AI & REF & AI & REF \\
\midrule
\endhead

\midrule
\multicolumn{8}{r@{}}{\small\itshape continued on next page} \\
\endfoot

\bottomrule
\endlastfoot

\multicolumn{8}{c}{\textbf{Bayesian nonparametrics}} \\
\midrule
\model{continuous_BART} & 1.0621 & 0.96 & 0.97 & 3749 & 4673 & 11.0 & 55.3 \\
\model{logistic_BART} & 1.0009 & 0.99 & --- & 5506 & --- & 3.5 & --- \\
\model{poisson_BART} & 1.0002 & 0.98 & --- & 23729 & --- & 13.7 & --- \\
\model{negative_binomial_BART} & 1.0004 & 0.96 & --- & 11538 & --- & 7.0 & --- \\
\model{probit_BART} & 1.0108 & 0.98 & 0.97 & 1956 & 1950 & 2.9 & 9.0 \\
\model{DPM_Gaussian} & 1.0027 & 0.95 & 0.94 & 495 & 419 & 4.4 & 529.4 \\
\model{GP_normal} & 1.0001 & 0.96 & 0.96 & 9524 & 11988 & 2.7 & 7.0 \\
\model{gp_regr} & 1.0001 & 0.95 & 0.94 & 17792 & 17060 & 3.0 & 8.5 \\
\model{gp_pois_regr} & 1.0001 & 0.95 & 0.95 & 16833 & 8862 & 25.9 & 33.3 \\
\model{hierarchical_gp} & 1.0004 & 0.94 & 0.96 & 2293 & 5616 & 12.9 & 62.2 \\
\model{kronecker_gp} & 1.0002 & 0.92 & 0.92 & 9205 & 11384 & 3.7 & 7.2 \\
\model{accel_gp} & 1.0038 & 0.66 & 0.67 & 803 & 778 & 31.2 & 261.8 \\
\model{nn_rbm1bJ10} & 1.0004 & 0.94 & 0.93 & 5039 & 10026 & 7.1 & 61.7 \\
\model{nn_rbm1bJ100} & 1.3085 & 0.51 & 0.94 & 3 & 11982 & 1447.7 & 530.2 \\
\midrule

\multicolumn{8}{c}{\textbf{Mixture / clustering}} \\
\midrule
\model{normal_mixture} & 1.0001 & 0.74 & 0.75 & 19217 & 18750 & 0.02 & 2.7 \\
\model{normal_mixture_k} & 1.0020 & 0.90 & 0.90 & 321 & 1652 & 1.0 & 110.1 \\
\model{low_dim_gauss_mix} & 1.0002 & 0.93 & 0.90 & 6624 & 6865 & 1.2 & 10.7 \\
\model{low_dim_gauss_mix_collapse} & 1.0002 & 0.93 & 0.92 & 5253 & 6554 & 1.3 & 8.5 \\
\midrule

\multicolumn{8}{c}{\textbf{Discrete latent variable models}} \\
\midrule
\model{Dirac_Spike_Laplace} & 1.0007 & 0.94 & 0.95 & 9142 & 1145 & 5.6 & 84.2 \\
\model{Dirac_Spike_normal} & 1.0006 & 0.97 & 0.96 & 9948 & 1405 & 7.3 & 48.7 \\
\model{ldaK2} & 1.0007 & 0.97 & 0.97 & 1385 & 8196 & 0.15 & 30.9 \\
\model{ldaK5} & 1.0003 & 0.96 & 0.96 & 4096 & 8265 & 0.22 & 138.3 \\
\model{poisson_changepoint} & 1.0002 & 0.97 & 0.95 & 8036 & 7002 & 0.01 & 3.9 \\
\model{poisson_k_changepoint} & 1.0002 & 0.95 & 0.96 & 5292 & 6734 & 0.05 & 8.9 \\
\model{iohmm_reg} & 1.0002 & 0.97 & 0.90 & 13668 & 12837 & 5.4 & 63.5 \\
\model{M0_model} & 1.0001 & 0.96 & 0.96 & 9112 & 11628 & 0.02 & 2.5 \\
\model{Mb_model} & 1.0012 & 0.94 & 0.95 & 2150 & 10152 & 0.02 & 10.6 \\
\model{Mh_model} & 1.0010 & 0.98 & 0.97 & 490 & 2823 & 0.29 & 8.3 \\
\model{Mt_model} & 1.0013 & 0.96 & 0.95 & 10463 & 20028 & 0.03 & 4.3 \\
\model{Mtbh_model} & 1.0010 & 0.96 & 0.97 & 575 & 3910 & 0.57 & 17.7 \\
\model{Mth_model} & 1.0180 & 0.98 & 0.99 & 434 & 3843 & 0.44 & 16.2 \\
\model{multi_occupancy} & 1.0098 & 0.95 & 0.95 & 267 & 229 & 0.58 & 111.2 \\
\model{Survey_model} & 1.0001 & 0.98 & 0.98 & 14214 & 3887 & 0.02 & 4.2 \\
\midrule

\multicolumn{8}{c}{\textbf{Hierarchical models}} \\
\midrule
\model{GLMM_Poisson_model} & 1.0016 & 0.97 & 0.97 & 1855 & 437 & 1.2 & 6.1 \\
\model{GLMM1_model} & 1.0003 & 0.96 & 0.96 & 4130 & 3843 & 0.08 & 1.9 \\
\model{bones_model} & 1.0003 & 0.95 & 0.95 & 10908 & 8220 & 0.60 & 141.4 \\
\model{dirichlet_meanprec_hier} & 1.0002 & 0.73 & 0.73 & 5756 & 12603 & 3.8 & 75.0 \\
\model{eight_schools_centered} & 1.0011 & 0.94 & 0.94 & 13294 & 665 & 0.02 & 5.9 \\
\model{eight_schools_noncentered} & 1.0002 & 0.94 & 0.94 & 7076 & 7593 & 0.12 & 1.4 \\
\model{election88_full} & 1.0025 & 0.97 & 0.96 & 1727 & 520 & 1.3 & 38.8 \\
\model{pilots} & 1.0233 & 0.97 & 0.96 & 13051 & 1492 & 0.09 & 59.8 \\
\model{rats_model} & 1.0003 & 0.95 & 0.95 & 6004 & 11304 & 0.02 & 3.2 \\
\model{surgical_model} & 1.0005 & 0.96 & 0.96 & 7890 & 1450 & 0.20 & 1.9 \\
\model{radon_county} & 1.0002 & 0.96 & 0.96 & 12412 & 15507 & 0.04 & 1.3 \\
\model{radon_county_intercept} & 1.0001 & 0.94 & 0.94 & 15564 & 12658 & 0.02 & 2.5 \\
\model{radon_hierarchical_intercept_centered} & 1.0003 & 0.95 & 0.95 & 7368 & 14294 & 0.04 & 4.9 \\
\model{radon_hierarchical_intercept_noncentered} & 1.0002 & 0.95 & 0.95 & 15643 & 5247 & 0.05 & 5.0 \\
\model{radon_partially_pooled_centered} & 1.0002 & 0.96 & 0.95 & 14782 & 15161 & 0.07 & 2.3 \\
\model{radon_partially_pooled_noncentered} & 1.0002 & 0.95 & 0.95 & 4773 & 4591 & 0.32 & 3.4 \\
\model{radon_variable_intercept_centered} & 1.0002 & 0.94 & 0.94 & 7470 & 15424 & 0.04 & 3.6 \\
\model{radon_variable_intercept_noncentered} & 1.0002 & 0.94 & 0.94 & 11024 & 5166 & 0.04 & 4.4 \\
\model{radon_variable_intercept_slope_centered} & 1.0006 & 0.94 & 0.94 & 12026 & 1915 & 0.12 & 8.5 \\
\model{radon_variable_intercept_slope_noncentered} & 1.0002 & 0.94 & 0.94 & 7639 & 5771 & 0.05 & 8.3 \\
\model{radon_variable_slope_centered} & 1.0002 & 0.95 & 0.95 & 13410 & 12532 & 0.02 & 4.1 \\
\model{radon_variable_slope_noncentered} & 1.0001 & 0.95 & 0.95 & 12461 & 5616 & 0.05 & 4.7 \\
\model{seeds_centered_model} & 1.0005 & 0.95 & 0.96 & 4556 & 4085 & 0.68 & 4.4 \\
\model{seeds_model} & 1.0003 & 0.95 & 0.94 & 4373 & 4637 & 9.1 & 5.0 \\
\model{seeds_stanified_model} & 1.0004 & 0.97 & 0.97 & 6114 & 3755 & 0.70 & 3.7 \\
\model{pl2_latent_reg_irt} & 1.0020 & 0.94 & 0.94 & 869 & 7090 & 4.8 & 24.0 \\
\model{gpcm_latent_reg_irt} & 1.0004 & 0.91 & 0.91 & 4173 & 3791 & 5.7 & 201.1 \\
\model{grsm_latent_reg_irt} & 1.0004 & 0.94 & 0.94 & 2978 & 4001 & 10.2 & 267.1 \\
\model{hier_2pl} & 1.0021 & 0.96 & 0.94 & 1556 & 641 & 22.1 & 71.9 \\
\model{irt_2pl} & 1.0047 & 0.98 & 0.96 & 411 & 265 & 6.8 & 39.9 \\
\model{lsat_model} & 1.0008 & 0.95 & 0.95 & 2384 & 5428 & 2.3 & 10.5 \\
\midrule

\multicolumn{8}{c}{\textbf{Time series / state-space}} \\
\midrule
\model{arK} & 1.0001 & 0.96 & 0.95 & 16337 & 8292 & 0.06 & 3.5 \\
\model{arma11} & 1.0010 & 0.97 & 0.89 & 1030 & 3291 & 0.10 & 4.1 \\
\model{garch11} & 1.0002 & 0.93 & 0.93 & 5285 & 9778 & 0.78 & 15.3 \\
\model{prophet} & 1.0004 & 0.92 & 0.90 & 7846 & 7471 & 4.0 & 12.7 \\
\model{state_space_stochastic_level_stochastic_seasonal} & 1.0047 & 0.95 & 0.95 & 133 & 281 & 0.99 & 24.1 \\
\midrule

\multicolumn{8}{c}{\textbf{ODE / Mechanistic models}} \\
\midrule
\model{sir} & 1.0004 & 0.96 & 0.96 & 3074 & 3478 & 6.1 & 404.3 \\
\model{covid19imperial_v2} & 1.0013 & 0.66 & 0.66 & 1661 & 1654 & 8.1 & 40.8 \\
\model{covid19imperial_v3} & 1.0046 & 0.64 & 0.59 & 3318 & 374 & 51.3 & 85.5 \\
\model{lotka_volterra} & 1.0002 & 0.95 & 0.94 & 5126 & 7871 & 33.8 & 188.4 \\
\model{one_comp_mm_elim_abs} & 1.0002 & 0.91 & 0.91 & 9560 & 9508 & 3.1 & 331.2 \\
\model{soil_incubation} & 1.0041 & 0.93 & 0.93 & 668 & 9260 & 2.2 & 1190.5 \\
\model{dugongs_model} & 1.0003 & 0.93 & 0.94 & 13267 & 5045 & 0.06 & 1.9 \\
\model{losscurve_sislob} & 1.0002 & 0.96 & 0.96 & 7893 & 7534 & 0.46 & 76.5 \\
\midrule

\multicolumn{8}{c}{\textbf{Graphical models}} \\
\midrule
\model{bym2_offset_only} & 1.0004 & 0.97 & 0.97 & 2909 & 2922 & 17.8 & 98.9 \\
\model{hmm_drive_0} & 1.0001 & 0.96 & 0.96 & 7335 & 21724 & 4.5 & 33.5 \\
\model{hmm_drive_1} & 1.0001 & 0.97 & 0.97 & 18917 & 16512 & 0.31 & 58.4 \\
\model{hmm_example} & 1.0001 & 0.94 & 0.95 & 19187 & 11133 & 0.10 & 17.5 \\
\model{hmm_gaussian} & 1.0001 & 0.97 & 0.85 & 14650 & 16889 & 0.81 & 77.5 \\
\model{hex_ising_seg_0} & 1.0018 & 0.94 & 0.93 & 7381 & 7496 & 1.3 & 7.5 \\
\model{pois_hmm_k3_0} & 1.0232 & 0.95 & 0.95 & 11829 & 16937 & 1.5 & 158.9 \\
\model{water_bif} & 1.0885 & 0.97 & 0.97 & 44 & 165 & 0.53 & 264.9 \\
\midrule

\multicolumn{8}{c}{\textbf{Regression models}} \\
\midrule
\model{blr} & 1.0001 & 0.95 & 0.95 & 15942 & 17705 & 0.07 & 0.79 \\
\model{diamonds} & 1.0001 & 0.96 & 0.96 & 15476 & 23738 & 0.08 & 0.79 \\
\model{earn_height} & 1.0011 & 0.93 & 0.93 & 18870 & 5376 & 0.00 & 5.4 \\
\model{kilpisjarvi} & 1.0018 & 0.94 & 0.94 & 18247 & 17333 & 0.01 & 0.77 \\
\model{radon_pooled} & 1.0001 & 0.97 & 0.97 & 18366 & 10809 & 0.06 & 1.6 \\
\model{kidscore_interaction} & 1.0001 & 0.95 & 0.94 & 16937 & 10681 & 0.03 & 1.6 \\
\model{kidscore_interaction_c} & 1.0019 & 0.93 & 0.93 & 16943 & 21589 & 0.02 & 0.99 \\
\model{kidscore_interaction_c2} & 1.0020 & 0.94 & 0.94 & 16942 & 13743 & 0.03 & 1.2 \\
\model{kidscore_interaction_z} & 1.0019 & 0.93 & 0.94 & 16942 & 21321 & 0.02 & 1.1 \\
\model{kidscore_mom_work} & 1.0020 & 0.93 & 0.94 & 16942 & 7417 & 0.02 & 1.4 \\
\model{kidscore_momhs} & 1.0001 & 0.94 & 0.95 & 18168 & 8401 & 0.05 & 1.2 \\
\model{kidscore_momhsiq} & 1.0001 & 0.97 & 0.97 & 19439 & 7306 & 0.02 & 3.1 \\
\model{kidscore_momiq} & 1.0001 & 0.96 & 0.96 & 19436 & 5810 & 0.02 & 2.2 \\
\model{mesquite} & 1.0011 & 0.56 & 0.56 & 16679 & 12238 & 0.04 & 7.3 \\
\model{logmesquite} & 1.0011 & 0.94 & 0.94 & 16679 & 9593 & 0.04 & 7.8 \\
\model{logmesquite_logva} & 1.0011 & 0.96 & 0.96 & 17987 & 8843 & 0.02 & 7.3 \\
\model{logmesquite_logvas} & 1.0012 & 0.95 & 0.95 & 16679 & 9644 & 0.04 & 13.0 \\
\model{logmesquite_logvash} & 1.0012 & 0.95 & 0.95 & 17164 & 9748 & 0.03 & 11.4 \\
\model{logmesquite_logvolume} & 1.0018 & 0.93 & 0.93 & 18350 & 9027 & 0.00 & 0.86 \\
\model{log10earn_height} & 1.0018 & 0.96 & 0.96 & 18314 & 5240 & 0.02 & 5.0 \\
\model{logearn_height} & 1.0017 & 0.96 & 0.96 & 18280 & 5324 & 0.02 & 4.5 \\
\model{logearn_height_male} & 1.0019 & 0.96 & 0.96 & 17540 & 7132 & 0.02 & 6.6 \\
\model{logearn_interaction} & 1.0020 & 0.96 & 0.96 & 16943 & 4575 & 0.02 & 14.6 \\
\model{logearn_interaction_z} & 1.0019 & 0.94 & 0.94 & 16943 & 11021 & 0.02 & 1.4 \\
\model{logearn_logheight_male} & 1.0018 & 0.95 & 0.95 & 17540 & 6381 & 0.02 & 20.1 \\
\model{nes} & 1.0018 & 0.94 & 0.94 & 14605 & 9328 & 0.04 & 3.4 \\
\model{sesame_one_pred_a} & 1.0022 & 0.93 & 0.93 & 17994 & 16611 & 0.00 & 0.66 \\
\model{GLM_Binomial_model} & 1.0001 & 0.94 & 0.94 & 18243 & 8583 & 0.13 & 1.2 \\
\model{dogs} & 1.0001 & 0.93 & 0.93 & 9554 & 8344 & 0.72 & 9.4 \\
\model{dogs_hierarchical} & 1.0001 & 0.93 & 0.94 & 17443 & 8451 & 0.15 & 2.6 \\
\model{dogs_log} & 1.0001 & 0.94 & 0.94 & 15240 & 5262 & 0.25 & 4.5 \\
\model{dogs_nonhierarchical} & 1.0025 & 0.99 & 0.99 & 10301 & 9521 & 2.2 & 16.8 \\
\model{nes_logit_model} & 1.0002 & 0.95 & 0.95 & 3686 & 8360 & 0.22 & 0.72 \\
\model{wells_daae_c_model} & 1.0002 & 0.88 & 0.88 & 13349 & 6084 & 1.4 & 10.2 \\
\model{wells_dae_c_model} & 1.0002 & 0.78 & 0.78 & 19018 & 6157 & 0.17 & 1.7 \\
\model{wells_dae_inter_model} & 1.0012 & 0.70 & 0.71 & 503 & 4944 & 1.0 & 2.4 \\
\model{wells_dae_model} & 1.0001 & 0.96 & 0.97 & 11648 & 9660 & 6.3 & 7.1 \\
\model{wells_dist} & 1.0001 & 0.91 & 0.91 & 15543 & 5142 & 0.25 & 1.0 \\
\model{wells_dist100_model} & 1.0001 & 0.95 & 0.95 & 11815 & 5369 & 0.66 & 0.99 \\
\model{wells_dist100ars_model} & 1.0007 & 0.84 & 0.85 & 735 & 6085 & 0.68 & 1.7 \\
\model{wells_interaction_c_model} & 1.0002 & 0.87 & 0.87 & 17933 & 7472 & 0.12 & 1.2 \\
\model{wells_interaction_model} & 1.0030 & 0.74 & 0.76 & 140 & 3431 & 0.67 & 2.2 \\
\model{Rate_1_model} & 1.0001 & 0.97 & 0.97 & 19082 & 6181 & 0.00 & 0.36 \\
\model{Rate_2_model} & 1.0001 & 0.95 & 0.95 & 19108 & 12852 & 0.06 & 0.47 \\
\model{Rate_3_model} & 1.0001 & 0.95 & 0.95 & 19564 & 6399 & 0.01 & 0.45 \\
\model{Rate_4_model} & 1.0001 & 0.97 & 0.97 & 19634 & 12314 & 0.00 & 0.47 \\
\model{Rate_5_model} & 1.0001 & 0.95 & 0.96 & 19288 & 6377 & 0.00 & 0.46 \\
\model{GLM_Poisson_model} & 1.0002 & 0.92 & 0.94 & 7388 & 6471 & 0.54 & 2.0 \\
\model{two_normal_spike_slab_model} & 1.0060 & 0.95 & 0.95 & 4510 & 12172 & 0.19 & 2.5 \\
\model{logistic_regression_rhs} & 1.0002 & 0.94 & 0.94 & 10558 & 9989 & 1.6 & 5.5 \\
\model{accel_splines} & 1.0012 & 0.93 & 0.93 & 1381 & 3154 & 7.2 & 26.4 \\
\end{longtable}
\endgroup

\newpage
\section{Evaluation on newly proposed Bayesian models}
\label{supp:new_models}
To assess whether AI4BayesCode generalizes beyond models that may have appeared in the training data of the code-generating LLM, we constructed additional models from recent arXiv Bayesian papers. Because the reported training-data cutoff of Claude Opus 4.7 (max effort) is January 2026, we selected ten arXiv papers whose first submissions appeared from February 2026 onward. Seven papers included public reference implementations, while three did not. For each paper, we provided only the natural-language Bayesian model specification and asked AI4BayesCode to generate a sampler without access to the paper, source code, or reference implementation. The generated samplers were evaluated using the same criteria as in Experiment~1. We summarize the results in Web Table~\ref{tab:sim1_arxiv_results}. AI4BayesCode successfully generated validated samplers for these newly proposed models, providing evidence that its performance is not solely attributable to memorization of existing implementations. Instead, the results suggested that the system generalizes by decomposing new model descriptions into modular sampling blocks and mapping them to built-in sampling components through AI Skills.

\begingroup
\fontsize{9.0}{10.5}\selectfont
\renewcommand{\arraystretch}{1.05}
\setlength{\tabcolsep}{3pt}

\setlength{\LTcapwidth}{\textwidth}
\captionsetup[longtable]{
    justification=raggedright,
    singlelinecheck=false
}

\begin{longtable}{@{}
    >{\raggedright\arraybackslash}p{4.9cm}
    >{\raggedright\arraybackslash}p{1.9cm}
    >{\centering\arraybackslash}p{1.25cm}
    >{\centering\arraybackslash}p{0.7cm}
    >{\centering\arraybackslash}p{0.7cm}
    >{\centering\arraybackslash}p{0.85cm}
    >{\centering\arraybackslash}p{0.85cm}
    >{\centering\arraybackslash}p{0.85cm}
    >{\centering\arraybackslash}p{0.85cm}
    @{}}

\caption{Results on the new arXiv paper benchmark. Each row corresponds to a Bayesian model described in a recent arXiv paper, cited in the second column. We summarize 100 replicate datasets per model by reporting the median of the maximum rank-normalized $\widehat R$ across parameters, the mean coverage of 95\% credible intervals across parameters, the median minimum bulk effective sample size, and the median running time. Models in the upper block have a reference implementation (REF), whereas those in the lower block do not and report ``---'' in the reference columns.}
\label{tab:sim1_arxiv_results} \\

\toprule
\multirow{2}{*}{\textbf{Paper}} &
\multirow{2}{*}{\textbf{Reference}} &
\multirow{2}{*}{\shortstack[c]{\textbf{Median}\\\textbf{max} $\boldsymbol{\widehat R}$}} &
\multicolumn{2}{c}{\textbf{95\% CI coverage}} &
\multicolumn{2}{c}{\textbf{Min bulk ESS}} &
\multicolumn{2}{c}{\textbf{Runtime (s)}} \\
\cmidrule(lr){4-5}
\cmidrule(lr){6-7}
\cmidrule(lr){8-9}
& & & AI & REF & AI & REF & AI & REF \\
\midrule
\endfirsthead

\multicolumn{9}{@{}l@{}}{\small\itshape
(Table~\thetable\ continued from previous page)} \\
\toprule
\multirow{2}{*}{\textbf{Paper}} &
\multirow{2}{*}{\textbf{Reference}} &
\multirow{2}{*}{\shortstack[c]{\textbf{Median}\\\textbf{max} $\boldsymbol{\widehat R}$}} &
\multicolumn{2}{c}{\textbf{95\% CI coverage}} &
\multicolumn{2}{c}{\textbf{Min bulk ESS}} &
\multicolumn{2}{c}{\textbf{Runtime (s)}} \\
\cmidrule(lr){4-5}
\cmidrule(lr){6-7}
\cmidrule(lr){8-9}
& & & AI & REF & AI & REF & AI & REF \\
\midrule
\endhead

\midrule
\multicolumn{9}{r@{}}{\small\itshape continued on next page} \\
\endfoot

\bottomrule
\endlastfoot

\multicolumn{9}{c}{\textbf{Models with reference implementations}} \\
\midrule

PliableBVS: A flexible Bayesian variable selection method for modeling interactions with mandatory modifying variables &
\citet{Asenso2026} &
1.0091 & 0.99 & 0.99 & 989 & 11256 & 21.4 & 57.8 \\
\cmidrule{1-1}

Bayesian Inference of Nonlinear Malaria Dynamics in Ghana via an Ensemble Markov Chain Monte Carlo Sampler &
\citet{AnsahNarh2026} &
1.0115 & 0.79 & 0.78 & 679 & 13496 & 42.0 & 18.2 \\
\cmidrule{1-1}

cyclinbayes: Bayesian Causal Discovery with Linear Non-Gaussian Directed Acyclic and Cyclic Graphical Models &
\citet{Lee2026} &
1.0222 & 0.86 & 0.86 & 2242 & 4954 & 2443.1 & 63.1 \\
\cmidrule{1-1}

A Bayesian Hierarchical Hurdle Beta-Binomial Model for Survey-Weighted Bounded Counts and Its Application to Childcare Enrollment &
\citet{Lee2026a} &
1.0032 & 0.90 & 0.91 & 5135 & 5882 & 1499.7 & 234.8 \\
\cmidrule{1-1}

Bayesian Inference for Non-Conjugate Distance Dependent Chinese Restaurant Process Models &
\citet{Marsh2026} &
1.0097 & 0.98 & 0.99 & 196 & 496 & 37.3 & 19.1 \\
\cmidrule{1-1}

Mixture-of-Finite-Mixtures Wishart Model for Clustering Covariance Matrices with an Application to Brain Functional Connectivity &
\citet{Li2026} &
1.0004 & 0.68 & 0.66 & 6756 & 3540 & 6.4 & 356.9 \\
\cmidrule{1-1}

Bayesian covariance regression for differential network analysis of zero-inflated microbiome data &
\citet{Xu2026} &
1.0024 & 0.96 & 0.89 & 4055 & 5815 & 185.2 & 1141.7 \\

\midrule
\multicolumn{9}{c}{\textbf{Models without reference implementations}} \\
\midrule

Bayesian change-plane regression &
\citet{Ohnishi2026} &
1.0021 & 0.97 & --- & 1191 & --- & 43.0 & --- \\
\cmidrule{1-1}

Bayesian Environment Invariant Regression &
\citet{Zhang2026} &
1.0010 & 0.82 & --- & 4219 & --- & 682.4 & --- \\
\cmidrule{1-1}

Bayesian inference of sparsity in stable vector autoregressive processes &
\citet{Heaps2026} &
1.0019 & 0.96 & --- & 940 & --- & 856.9 & --- \\

\end{longtable}
\endgroup

\section{Kernel-tier Dependency}
Web Table~\ref{tab:kernel-vendors} lists the external numerical libraries used in the kernel tier of AI4BayesCode and their licenses.

\begin{table}[h]
\centering
\begin{tabular}{lll}
\hline
\textbf{Library} & \textbf{Version/snapshot} & \textbf{License} \\
\hline
\texttt{BART}      & 2.9.10      & GPL-2.0-or-later \\
\texttt{SoftBART}  & 1.0.3      & GPL-2.0-or-later \\
\texttt{mcmclib}   & accessed April 2026 & Apache-2.0 \\
\texttt{celerite}  & 0.4.0      & MIT \\
\texttt{libgp}     & 0.3.0      & BSD-3-Clause \\
\texttt{librjmcmc} & pre-2012 snapshot & CeCILL-B \\
\hline
\end{tabular}
\caption{External numerical libraries used in the kernel tier of AI4BayesCode.}
\label{tab:kernel-vendors}
\end{table}

  \bibliography{bibliography.bib}